\documentclass[fleqn,usenatbib]{mnras}

\usepackage{newtxtext,newtxmath}

\usepackage[T1]{fontenc}
\usepackage{ae,aecompl}

\usepackage{graphicx}	
\usepackage{amsmath}	
\usepackage{amssymb}	

\title[Fragmentation of debris streams from TDEs]{What causes the fragmentation of debris streams in TDEs?}

\author[Sacchi et al.]{
Andrea Sacchi$^{1,2}$\thanks{E-mail: sacchi@arcetri.astro.it},
Giuseppe Lodato$^{2}$,
Claudia Toci$^{3}$,
Valentina Motta$^{2}$
\\
$^{1}$Dipartimento di Fisica e Astronomia, Universit\`a degli Studi di Firenze, Via G. Sansone 1, I-50019, Sesto Fiorentino, Italy\\
$^{2}$Dipartimento di Fisica, Universit\`a degli Studi di Milano, Via Celoria 16, 20133, Milano, Italy\\
$^{3}$INAF OA Brera, Via Brera, 28, 20121, Milano, Italy\\
}

\date{Accepted XXX. Received YYY; in original form ZZZ}

\pubyear{2019}

\begin{document}
\label{firstpage}
\pagerange{\pageref{firstpage}--\pageref{lastpage}}
\maketitle

\begin{abstract}
A tidal disruption event (TDE) occurs when a star passes too close to a supermassive black hole and gets torn apart by its gravitational tidal field. After the disruption, the stellar debris form an expanding gaseous stream. The morphology and evolution of this stream is particularly interesting as it ultimately determines the observational properties of the event itself.
In this work we perform 3D hydrodynamical simulations of the TDE of a star modelled as a polytropic sphere of index $\gamma=5/3$, and study the gravitational stability of the resulting gas stream. We provide an analytical solution for the evolution of the stream in the bound, unbound and marginally bound case, that allows us to describe the stream properties and analyse the time-scales of the physical processes involved, applying a formalism developed in star formation context.
Our results are that, when fragmentation occurs, it is fueled by the failure of pressure in supporting the gas against its self-gravity. 
We also show that a stability criterion that includes also the stream gas pressure proves to be far more accurate than one that only considers the black hole tidal forces, giving analytical predictions of the time evolution of the various forces associated to the stream. Our results point out that fragmentation occurs on timescales longer compared with the observational windows of these events and is thus not expected to give rise to significant observational features.
\end{abstract}




\section{Introduction}\label{sec:introduction}
In the last two decades a new class of X-ray outbursts have started to be detected \citep{bad96,gez03}. Having high variability (with timescales ranging from weeks to even a few days) and bright but soft spectrum (L $\geq 10^{42 \text{-} 43}$ erg/s) along with other unusual characteristics, such as no sign of Seyfert activity of the host galaxy, the explanation for these events has to be sought in the then uncharted class of phenomena of tidal disruption events (TDEs).

These events happen when the strong gravity of a supermassive black hole (SMBH) disrupts a passing star through tidal forces, if the latter approaches the compact object at a distance smaller than the SMBH's tidal radius. The tidal radius is the distance at which the tidal force equals the stellar self-gravity and it is defined as $r_\textup{t}\approx r_\star(M_\textup{h}/M_\star)^{1/3}$, where $r_\star$ and $M_\star$ are the radius and mass of the star and $M_\textup{h}$ is the mass of the SMBH. 

Several putative TDEs have been found, observed in almost every band of the electromagnetic spectrum: soft X-ray \citep{bad96}, optical and UV \citep{gez12,kom12,kom15,hun17}, radio \citep{zau11}, hard X-ray and gamma \citep{blo11,cen12,bro15,auc17,bla17}. However, TDEs had been studied for almost twenty years before the observational discovery, both from a theoretical point of view \citep{lac82,ree88, phi89} and through numerical simulations \citep{lum82,lum83,eva89}, due to their importance as a tool to study the properties of BHs, especially in the center of galaxies.

In the simplest theoretical scenario, a star in hydrostatic equilibrium is set on a parabolic orbit around a SMBH, with pericenter distance equal to the tidal radius. Under the impulse approximation \citep{ree88}, meaning that the interaction between the two objects occurs instantaneously rather than gradually as the star approaches the SMBH, the former remains unperturbed up to the pericenter, where it is tidally disrupted. After that, roughly half of the stellar material is launched onto hyperbolic orbits allowing the debris to escape the system, while the other half remains bound to the SMBH in highly elliptical orbits and therefore will eventually return to the compact object forming a bright accretion disc. The most bound material is the first to complete its orbit and accrete onto the black hole. The timescale of this process is $t_\textup{min} \approx 41\,\textup{days}\:(M_\star/M_\odot)^{-1/2}(R_\star/R_\odot)^{3/2}(q/10^6)^{1/2}$, where $q = M_\textup{h}/M_\star$.

Under the assumption that the time the debris need to form the disc and accrete onto the black hole is much shorter than the time it takes to complete the elliptical orbit, the luminosity of the event is found to be proportional to the mass return rate at the pericentre:
\begin{equation}
    L\propto\frac{{\rm d}M}{{\rm d}t}=\frac{(2\upi GM_\textup{h})^{2/3}}{3}\frac{{\rm d}M}{{\rm d}E}t^{-5/3}.
\end{equation}
Assuming the flatness of the energy distribution, it is possible to find what is considered to be the ``smoking gun'' of these phenomena: a light curve that should fall as $t^{-5/3}$. Results from numerical simulations and analytical considerations, however, found that these phenomena depend on a wide range of parameters, such as the stellar internal structure \citep{lod09} and spin \citep{gol19,kag19,sac19}, the properties of the black hole \citep{hay16,tej17}, the penetration factor $\beta=r_\textup{t}/r_\textup{p}$ \citep{gui13} and the physics of disc formation \citep{hay13,bon16}, that can consistently modify the $t^{-5/3}$ behavior.

Furthermore, the TDEs' observable properties are closely connected to the stream of debris that forms after the stellar disruption. In particular, the self-gravity of the stream can become dominant in the transverse direction \citep{koc94} and it is therefore possible to encounter a gravitational instability which my lead to fragmentation. Initially studied in the pioneer works of \citet{cou15}, an instability criterion for a TDE debris stream can be obtained by defining a \textit{critical density} for the SMBH, $\rho_ \textup h \sim {M_\textup h}/{r^3},$ where $r$ is the distance of a debris fluid element from the SMBH, 
yielding a instability condition in the form $\rho > \rho_{\rm h}$, picturing a scenario where the critical density falls at a rate that is steeper than the debris density with respect to radial distance from the SMBH, which physically indicates that the rate at which the material is torn apart is slower than the rate at which it aggregates. In this condition the debris stream is subject to fragmentation and is therefore considered to be \textit{gravitationally unstable}.

If the debris is described by a polytropic model \citep{cha39}, the pressure $p$ is
$p = k \rho^{\gamma}$, where $k$ is a proportionality constant and $\gamma$ is called the polytropic index. Numerical simulations confirmed that the evolution of the debris stream and its density depend on $\gamma$ \citep{cou15}. 

Furthermore, \citet{cou16a,cou16b} showed that the stability of the streams depends on the polytropic index, and in particular revealed how, for $\gamma\gtrsim5/3$, the stream is susceptible to fragmentation. The newly formed clumps of material can strongly affect the light curve of the event, on account of the fact that the debris is not accreted ``continuously'' but rather in almost discrete blocks.

It is not clear whether a star with critical index $\gamma=5/3$ should be gravitationally stable or should fragment. The conclusion from \citet{cou16b} is that the star should be unstable for fragmentation, though only at later times: this is due to the fact that in this peculiar case the over-densities grow as a power-law rather than exponentially as in more compact ($\gamma>5/3$) cases. In this paper we wish to verify this result and better understand this pivotal case's behavior along with the stability criterion. In order to do so we will perform numerical simulations and analytical calculation, often considering the debris stream as in free-fall onto the black hole. This approximation, already employed by \citet{cou19} proved to be extremely effective in describing the behaviour of the marginally bound part of the stream.

The actual lightcurve of TDEs will depend on several other physical effects. The internal structure of the star may modify the long term evolution of the stream \citep{lod09,gui13,law19,ryu20}. The possible influence of a secondary black hole \citep{liu09,Coughlin17,Coughlin18,Vigneron18,Coughlinbinary} will affect the orbital evolution of the debris. Also, the incoming debris will be affected by relativistic precession \citep{hay13,bon16,Liptai19,Gafton19} and the emerging emission will certainly depend on the specific heating and cooling process of the accretion flow. Here, we concentrate specifically on the dynamics of the stream in the simplest configuration of a single black hole. The internal structure of the star, while modifying the long-term evolution is not expected to alter significantly the stability of the stream. 

The paper is organized as follows: in section 2 we will describe the numerical setup of the simulations we performed in order to test the $\gamma=5/3$ case: a ``standard" simulation without stellar rotation and a case with an initial stellar rotation. Our initial goal was to widen our understanding of the problem considering also the effect of rotation; in section 3 we will show the results of our simulations; in section 4 we will discuss the parallelism between the debris streams that form after a TDE and the gas filaments found in star forming regions and we will derive interesting analytical results using the formalism developed in that context;  finally in section 5 we will draw our conclusions.

\section{Numerical simulations}

In order to better understand and possibly clarify the nature of the fragmentation we performed numerical simulations. We suppose that the stream follows a polytropic equation of state $p=k\rho^\gamma$, where we recall that $p$ is the pressure of the gas, $\rho$ its density, $k$ is the polytropic constant and $\gamma$ the polytropic index, equal in our case to $5/3$. The simulations are performed using a polytropic sphere with adiabatic index $\gamma=5/3$ to model our to-be-disrupted star.
This value of polytropic index is also particularly interesting as it is the most common choice in all of the simulation in literature since \citet{nol82}.

The simulations employ a 3D Smoothed Particle Hydrodynamic (SPH) code, PHANTOM \citep{pri17}. SPH codes are particularly suitable for simulating TDEs as the majority of the space covered by our simulations is empty and these codes have the key property of linking the resolution of the simulation to the mass distribution.

The mass of the star and its radius are set to be $1\,M_\odot$ and $1\,R_\odot$, these are also chosen as our code units. Likewise the density unit is $\rho_0=M_\odot/R_\odot^3\approx5.09$ g/cm$^3$ and the time unit is $t_0=\sqrt{R_\odot^3/GM_\odot}\approx1590$ s, although more often we will adopt as time unit the minimum return time $t_\textup{min}$, that is the time it takes for the most bound debris of the disrupted star to complete an orbit around the black hole and come back to the pericenter. Considering, as it is the case of our simulations, a solar-like star disrupted by a SMBH of $10^6\,M_\odot$, $t_\textup{min}\approx41$ days \citep{ree88}.

In order to decide the optimal number of particles to use in our simulations we performed a convergence test. Convergence is reached at $\approx1.5\times10^6$ SPH particles. This number is higher than the one usually used to simulate TDEs \citep{eva89, aya00, bog04}, as we need to resolve the instability of the stream rather than its bulk motion.

\subsection{Numerical setup}
The polytropic sphere is initially relaxed without the black hole gravitational potential. A velocity damping is added in this phase in order to remove possible noise in the initial random displacement of the particles. 
After this first relaxation, we checked that the density profile of the sphere was the expected polytropic sphere, as described in \citet{lod09}.

In the case of a rotating star, after this first phase a rigid rotation is imposed and the sphere is relaxed once more, this time without the velocity damping (that would slow down the stellar rotation), similarly to \citealt{sac19}. The amount of rotation is described by the dimensionless parameter $\alpha=\omega/\omega_\textup{b}$, where $\omega$ is the stellar angular velocity and $\omega_\textup{b}$ is the break-up velocity, defined as the velocity at which the centrifugal force equals the stellar self-gravity. The non-rotating case corresponds to $\alpha=0$, while $\alpha=1$ indicates the maximally rotating case. The star is relaxed until it reaches a new equilibrium state that we monitor through the central density and thermal energy. Usually the value of $\alpha$ varies during the relaxation process \citep{sac19}.
However, the value of $\alpha$ in our simulation is relatively small, $\alpha=0.2$. For this slow value of stellar rotation a negligible reduction is observed during this second relaxation phase.

Once the star is relaxed, whether rotating or not, it is injected into a parabolic orbit around a $10^6\,M_\odot$  black hole, which is modelled as a Keplerian potential. The star is injected from a distance equal to 3 $r_\textup{t}$, with penetration factor $\beta=r_\textup{t}/r_\textup{p}=1$, where $\beta$ is the ratio between the tidal radius and pericentre distance $r_\textup{p}$. The penetration factor indicates how close to the hole the star passes during its orbit. Values of $\beta$ greater than 1 mean that the star undergoes a deep plunge in the black hole potential well, while values smaller than 1 mean a far away passage. A passage with $\beta=1$ guarantees a complete stellar disruption, for $\beta<1$ instead one only gets partial disruption or the stripping of stellar material \citep{gui13}.

After the star gets past the pericenter, it is disrupted. When it is sufficiently far from the source of the Keplerian potential, the latter is replaced with a sink particle of mass $10^6\,M_\odot$ and a relatively large accretion radius of 5 tidal radii. This means that every SPH particle that gets closer than 5 tidal radii from the black hole is considered to be accreted and removed from the simulation. This of course does not allow us to observe the physics of stream-stream shocks, circularization and accretion, however this choice is due to the fact that we are focusing our interest on the stream evolution properties.

\section{Results}

Here we illustrate the results of our simulations. We performed two sets of simulations, the first with a non-rotating star ($\alpha=0$) and a second set including the rotation of the star ($\alpha\neq0$ and in our particular case $\alpha=0.2$). To better highlight the major steps we divided the section as follows: we will first discuss how we determined the presence or absence of fragmentation in our simulation, then we will discuss the mechanism that generates it through time scale comparison and lastly we will show what happens if one adds an initial stellar rotation to the system.

\subsection{Fragmentation and convergence}

The first crucial result obtained through our simulation is that the stream of gas generated by the tidal disruption of a non-rotating star modelled as a polytropic sphere with adiabatic index $\gamma=5/3$ is affected by gravitational instabilities that bring it to fragment into smaller almost spherical blobs, as already shown by \citet{cou15}.

The presence of fragmentation is assessed via visual analysis of the stream appearance and through the analysis of the mean density 
and the density fluctuations, evaluated as the standard deviation of the mean density.

\begin{figure}
	\includegraphics[width=\columnwidth]{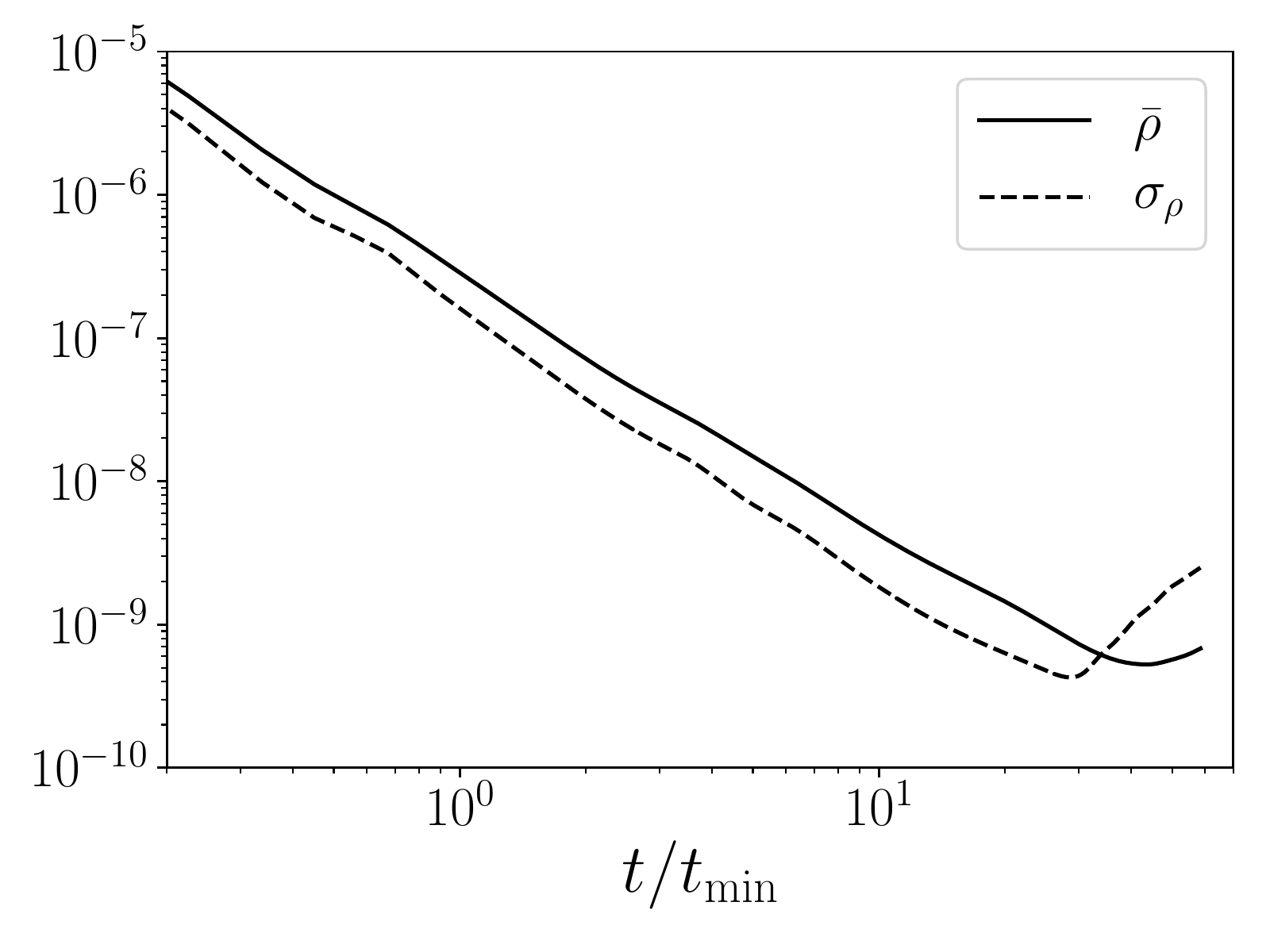}
    \caption{The solid line represent the mean density of the stream while the dashed one the density fluctuations, as functions of time, normalised to the minimum return time.}
    \label{fig:den_mean_fluc}
\end{figure}

Figure \ref{fig:den_mean_fluc} shows how initially both the mean value of the density (solid black line) and the fluctuations (dashed black line) fall as a power-law with power index $n\approx-1.67$. After almost 30 $t_\textup{min}$ the density fluctuations reach their minimum and start growing, soon overcoming the mean density of the stream. The turning point of the fluctuations is interpreted, in analogy with Cosmology \citep{pee80}, as the moment where fragmentation starts.

\begin{figure}
	\includegraphics[width=\columnwidth]{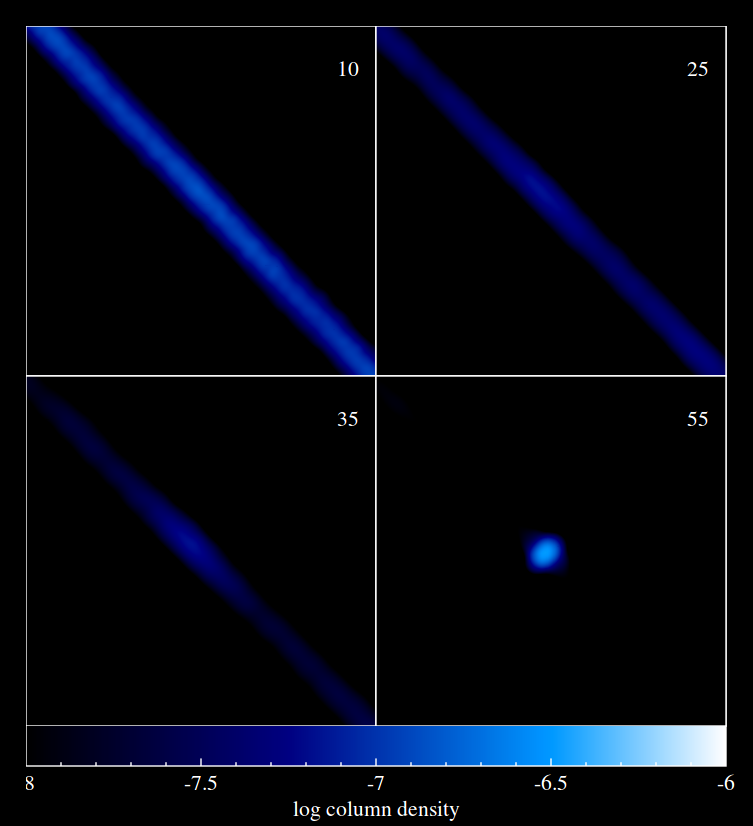}
    \caption{Projected density of the stream. Each panel shows the simulation taken at different times: at $t=$10, 25, 35 and 55 $t_\textup{min}$ from top left to bottom right.}
    \label{fig:den_rend}
\end{figure}

In order to confirm this technique we analyse also the projected density of the stream taking snapshots or our simulation. Figure \ref{fig:den_rend} shows a section of the stream at four different moments: $t=$10, 25, 35 and 55 $t_\textup{min}$ (that is as 13 months, almost 3 years, 4 years and 6 years after the stellar disruption). These moments corresponds to an instant way before the fragmentation, right before, right after and far after the fragmentation occurred. From these snapshots it is clear that the fragmentation occurs indeed when the density fluctuations reach the turning point.

\begin{figure*}
	\includegraphics[width=\textwidth]{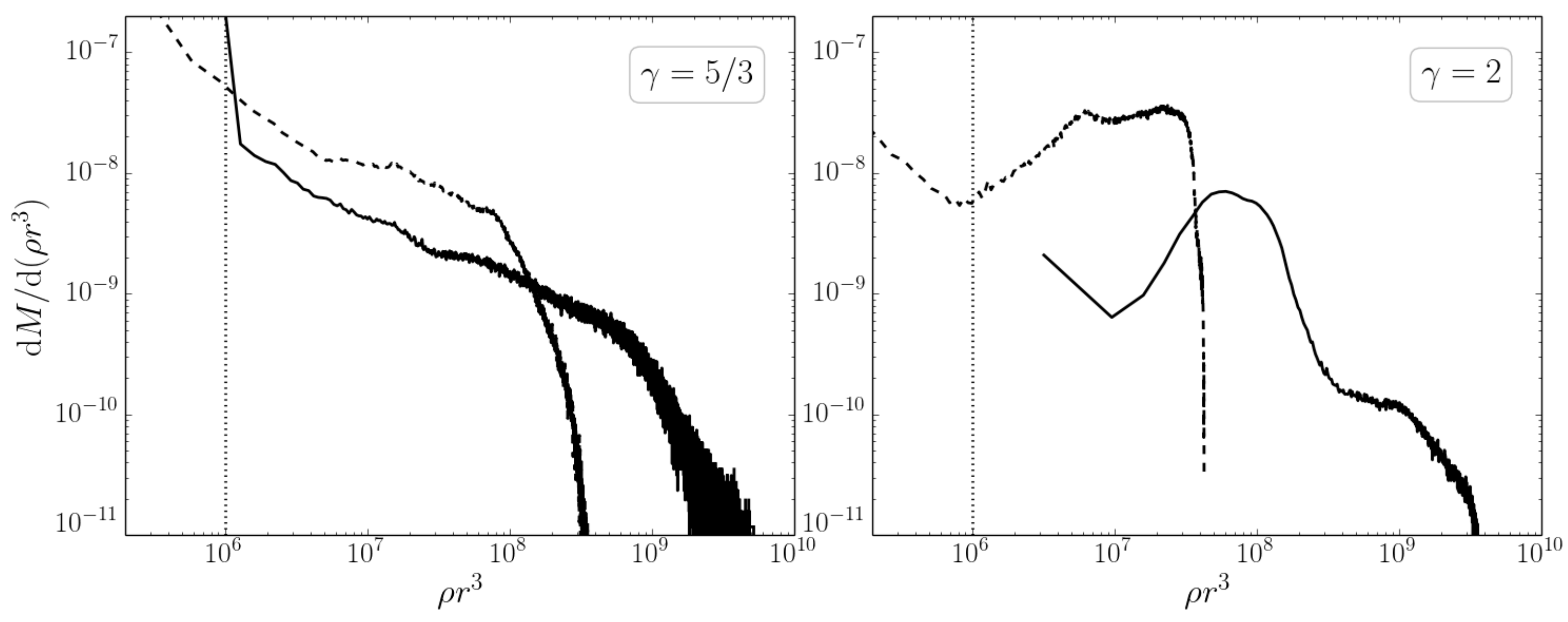}
    \caption{A representation of the tidal criterion for two different stellar structures (the primary case for this work alongside a more compact one with $\gamma = 2$) showing the ${\rm d}M/{\rm d}(\rho r^3)$ as a function of $\rho r^3$ at different times in the simulation, such as far before the fragmentation point ($t=0.3\,t_\textup{frag}$, dashed line) and far after ($t=2\,t_\textup{frag}$, solid line). In order for the tidal criterion to be satisfied $\rho r^3$ must be greater then $M_\textup{h}$, indicated by the vertical dashed line.}
    \label{fig:cou_crit}
\end{figure*}

The search for a criterion able to identify effectively the time at which the stream fragments is of particular interest as the one presented in the Introduction, suggested by \citet{cou16b}, does not give not an overly accurate estimate.
To be more precise, fig. \ref{fig:cou_crit} shows the mass distribution  ${\rm d}M/{\rm d}(\rho r^3)$ as a function of $\rho r^3$ for two stellar structures: the so far studied $\gamma = 5/3$ and a more compact case $\gamma = 2$, more susceptible to fragmentation. The distribution is shown at two times: when the stream is still far before the fragmentation point ($t=0.3\,t_\textup{frag}$, dashed line) and far after ($t=2\,t_\textup{frag}$, solid line).
Fig. \ref{fig:cou_crit} shows how the criterion based on the tides, $\rho r^3$ be greater than the black hole mass $M_\textup{h}=10^6\,M_\odot$ (indicated by the vertical dotted line in the figure), is satisfied by almost all of the stream, eccept for a small fraction of it ($\lesssim10\%$) already at the earliest time. The fact that the more compact scenario shows a distinct plateau at high densities after disruption might imply that a sharper criterion should have its critical point further up in the $\rho r^3$ ladder, instead of setting it at the black hole mass.

As suggested by \citet{cou16b}, based on simulations described in \citet{cou15} and \citet{cou16a}, the noise due to the numerical methods affects the time at which fragmentation occurs. The aforementioned convergence test has been performed in order to prove that the fragmentation is indeed physical and not only a result of numerical features of the code. The convergence has been considered satisfied when increasing the number of SPH particles employed in the simulation, the time at which fragmentation occurs would not increase appreciably. 

\begin{figure}
	\includegraphics[width=\columnwidth]{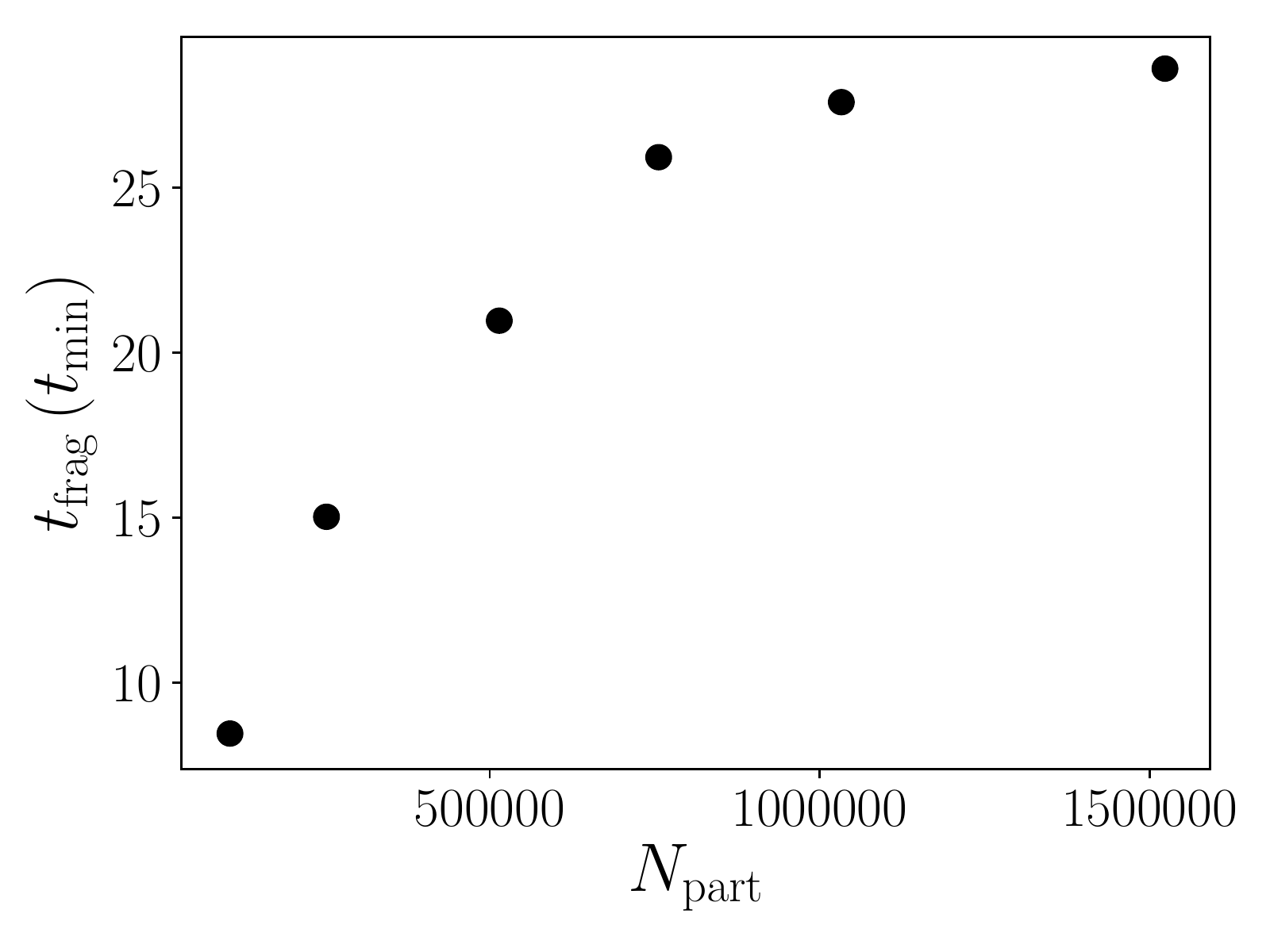}
    \caption{The figure shows the time at which the fragmentation occurs as a function of the number of SPH particles used in the simulation.}
    \label{fig:conver}
\end{figure}

Figure \ref{fig:conver} shows the time at which fragmentation occurs (hereafter $t_\textup{frag}$) as a function of the number of SPH particle of our simulations. When the simulation is performed using a number of particles smaller than $\approx10^6$, $t_\textup{frag}$ is strongly dependent on the resolution of the simulation, while above one million SPH particles it can be considered roughly constant. Note that the time of fragmentation occurs when the luminosity has already dropped by $\sim (30)^{-5/3}\sim 3\times 10^{-3}$ from the peak. We consider our choice of using, as mentioned in the previous section ($1.5\times10^6$ particles), satisfactory.

\subsection{Time scales}

We now consider the mechanisms responsible for the fragmentation. We compare the various contributions of the forces playing an active role in our picture by comparing their characteristic time scales. 

Representing the stream as a cylindrical fluid is a good assumption for the dynamics of these events, as also pointed out by \citealt{cou19}. In their work, the authors wrote the Lagrangian describing the motion of the core of the stream considering the effects of the black hole and the core mass gravitational forces, focusing on the fallback rate temporal behaviour and the bound core fate.

In this work, we are interested in studying the \textit{fluid} behaviour of the debris stream, its equilibrium state and its (eventual) fragmentation. 
This allow us to derive the temporal evolution of the stream quantities using simple physical considerations, introducing time scales related to the main forces at play and discussing their evolution with time.
Four are the forces that are actively involved in determining the dynamics of a fluid stream. Apart from the self-gravity of the stream and the tidal effects of the black hole, the internal forces of the gas (i.e. its pressure) and the background stretching of the gas are also at play. Indeed, the stream is affected by a stretching in the volume along the radial direction of the black hole, due to the differential acceleration of the particles composing the stream. This phenomenon is very important because it affects \textit{all} the quantities, introducing a time-dependency that becomes important as time passes by.

Each of these forces corresponds to a typical time scale: the stream self-gravity is linked to the free-fall time $t_\textup{ff}$, the tidal force generated by the black hole corresponds to a typical dynamical time $t_\textup{d}$, the gas pressure of the stream si associated to the sound crossing time $t_\textup{sc}$ and finally we have a stretching time scale $t_\textup{s}$.
As pointed out, we are interested in the debris stream: in the following analysis all these quantities are always referred to and computed for the part of the stream with positive energy.

The first force we consider is the stream self-gravity. This force will obviously favor the fragmentation and gravitational collapse and thus the fragmentation of the debris. The free-fall time is given by \citep{jea02}:
\begin{equation}
    t_\textup{ff} = \frac1{\sqrt{2\upi G\rho}},
\end{equation}
where $\rho$ is the density of the stream and the $2\upi$ factor accounts for the cylindrical geometry we are working with.

The tidal forces of the black hole tend to prevent any gravitational collapse. The dynamical time is computed as
\begin{equation}
    t_\textup{d} \simeq \frac1{\sqrt{ G\rho_\textup{h}}},
\end{equation}
where $\rho_\textup{h}=3M_\textup{h}/4\upi r^3$ is the ``density" of the black hole over a sphere of radius $r$.

\begin{figure}
	\includegraphics[width=\columnwidth]{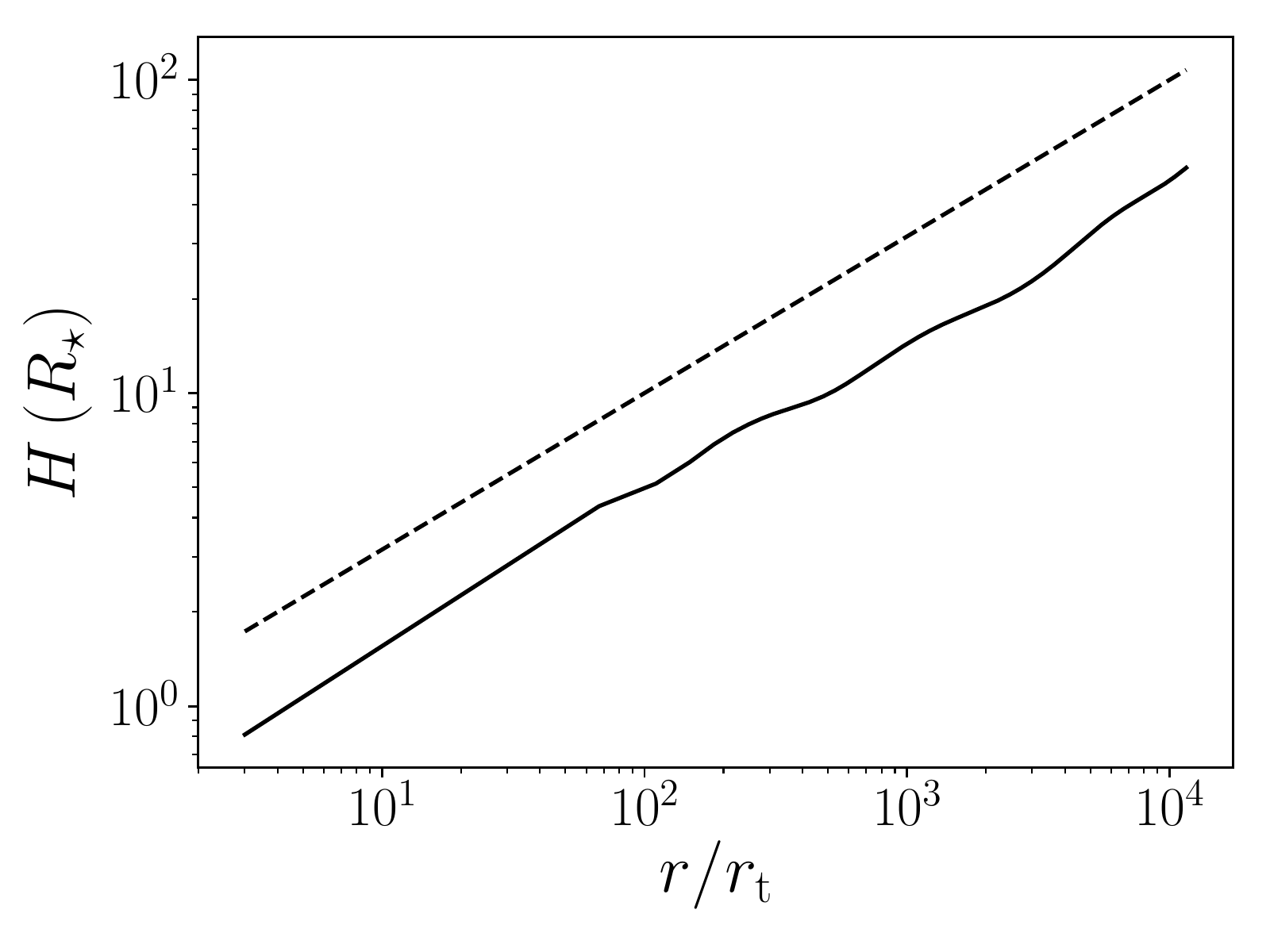}
    \caption{The solid black line shows the stream width as a function of the distance from the black hole normalized to the tidal radius. The dashed black line is the $r^{1/2}$ behaviour.}
    \label{fig:rad_dim}
\end{figure}

Pressure tends to stabilize the stream, the relative thermal time is just the sound crossing time across the transverse direction
\begin{equation}
    t_\textup{p}=\frac H{c_\textup{s}},
\end{equation}
where $H$ is the transverse size of the stream and $c_\textup{s}$ is the sound speed of the gas. The sound speed is calculated from the density, knowing the polytropic equation of state that describes the gas :
\begin{equation}
    c_\textup{s}=\sqrt{\frac{\partial p}{\partial\rho}\bigg|_{s}}=\sqrt{k\gamma\rho^{\gamma-1}},
\end{equation}
The stream width $H$ is obtained by averaging the distance of the particles composing the stream, considered again as a cylinder, from its axis: operatively we ``sliced" the stream along its length and for each slice we computed the mean distance of the particles within the slice from the cylinder axis, finally we mediated all the radii obtained along the stream. Figure \ref{fig:rad_dim} shows $H$ as a function of the distance from the black hole normalized to the tidal radius of the original star (solid black line). The dashed black line shows the $r^{1/2}$ behaviour found by \citet{cou16b} in disagreement with the $r^{1/4}$ prediction of \cite{koc94}.

The last time scale is associated to the stream stretching along its axis. This apparent force is due to the fact that right after the stellar disruption each gas particle lies on a different orbit determined by its energy. The spread in energy of the debris is given by the change of the black hole gravitational potential across the star at the time of disruption, that is at the first pericenter passage, in the impulse approximation \citep{lac82,ree88}. 
We suppose that the length $l$ of the stream changes with time according to the prescription $l=l_0a(t)$, where $l_0$ is the length of the stream at the initial time and $a(t)$ represents a dimensionless parametre of the stretching. 
This stretching introduces a proper timescale, defined in analogy with the Hubble time in Cosmology \citep{pee80} as
\begin{equation}
    t_\textup{s}=\frac a{\dot a}\label{for:str_time},
\end{equation}
where the dot indicates the time derivative.
The idea to consider the expansion and the contraction of the fluid volume in analogy with Cosmology using dimensionless parameters and an appropriate coordinate system has been already used to study the Solar wind fluctuations \citep{gra96,lan14,del15} as well as the collapse of pre-stellar cores \citep{toc18}.

\begin{figure}
	\includegraphics[width=\columnwidth]{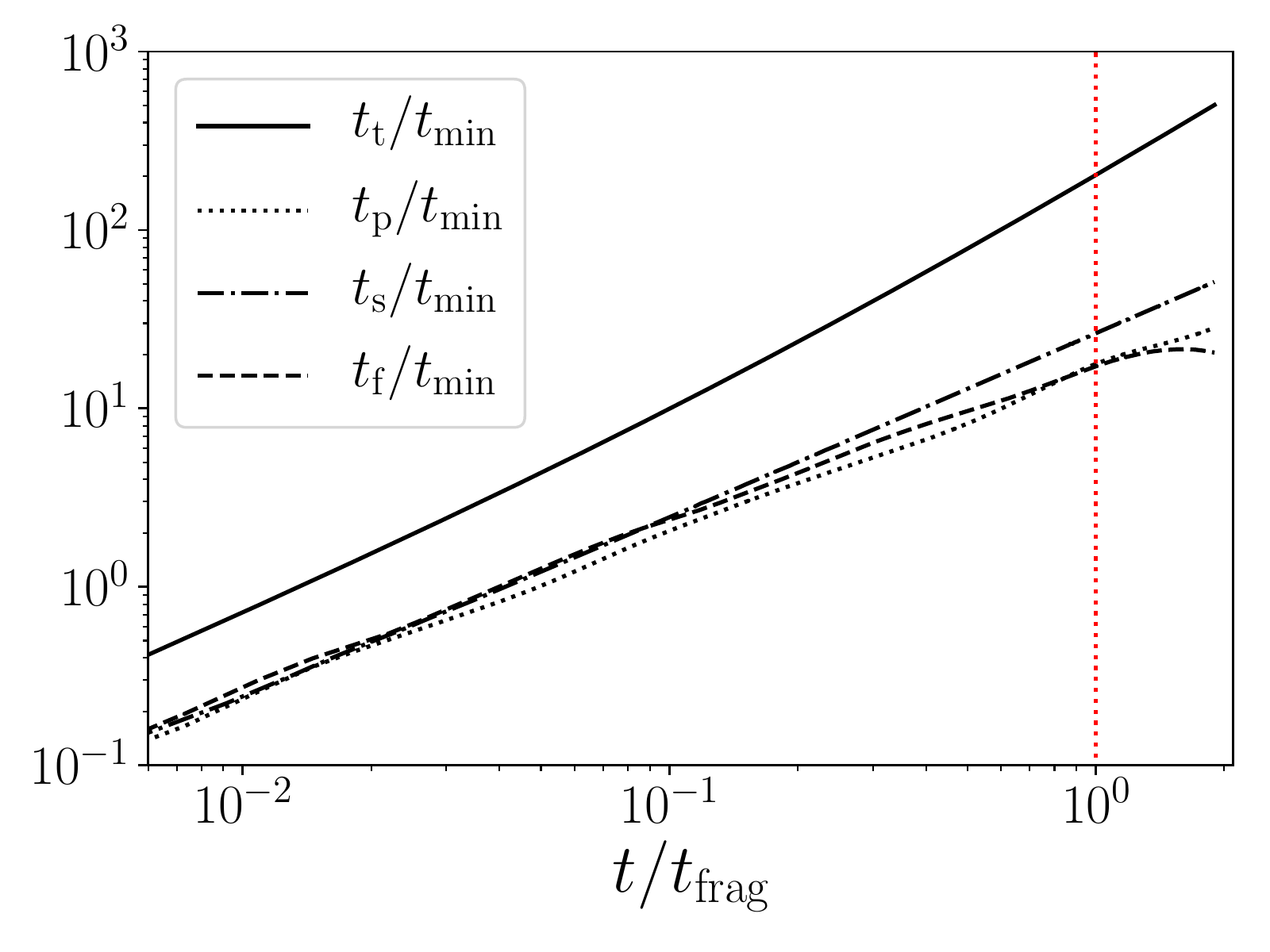}
    \caption{The figure shows the four described time scales: tidal (solid black line), thermal (dotted), stretching (dashed) and free-fall (dash-dotted). All the time scales are normalized to the minimum return time ($t_\textup{min}\approx41$d), while the time on $x$-axis is normalized to the time at which fragmentation occurs $t_\textup{frag}\approx30\,t_\textup{min}$.}
    \label{fig:tim_sca}
\end{figure}

Figure \ref{fig:tim_sca} shows the behaviour of the four described time scales as a function of time. The solid black line represents the tidal time, the dashed one the free-fall time, the dotted line is the thermal time and dash-dotted one is the stretching time. They are all normalized to the minimum return time (the time at which the first of the stellar debris comes back to its orbital pericenter, in our case $\approx41$ days) while the time on the $x$-axis is normalized to the fragmentation time $t_\textup{frag}\approx30\,t_\textup{min}$. The vertical dotted red line has been drawn to facilitate the comparison of the time scales at the fragmentation time.

The first consideration one can draw from figure \ref{fig:tim_sca} is that all the time scales have comparable magnitudes (within a factor $1.1$), crossing several times during the stream evolution. The only exception is the tidal time that is a factor $2.5-12$ longer than the others, although at the beginning of the simulation, this time scale should have been the smallest one in order for the star to be tidally disrupted.
This immediately shows that tidal forces are not responsible for fragmentation, where it happens.

Apart from the tidal time scale, all of the other time scales share a very similar evolution. The thermal time is the smallest one along almost the entirety, hence pressure is the most relevant force of the stream evolution, and it prevents its fragmentation. 
The time scales that disentangles from the others is the stretching time, that accelerates its growth after roughly $4\,t_\textup{min}$.
When fragmentation occurs the stretching time is almost a factor 2 larger then both the free-fall and thermal time, although it is still a 10 smaller then the tidal time.

\begin{figure}
	\includegraphics[width=\columnwidth]{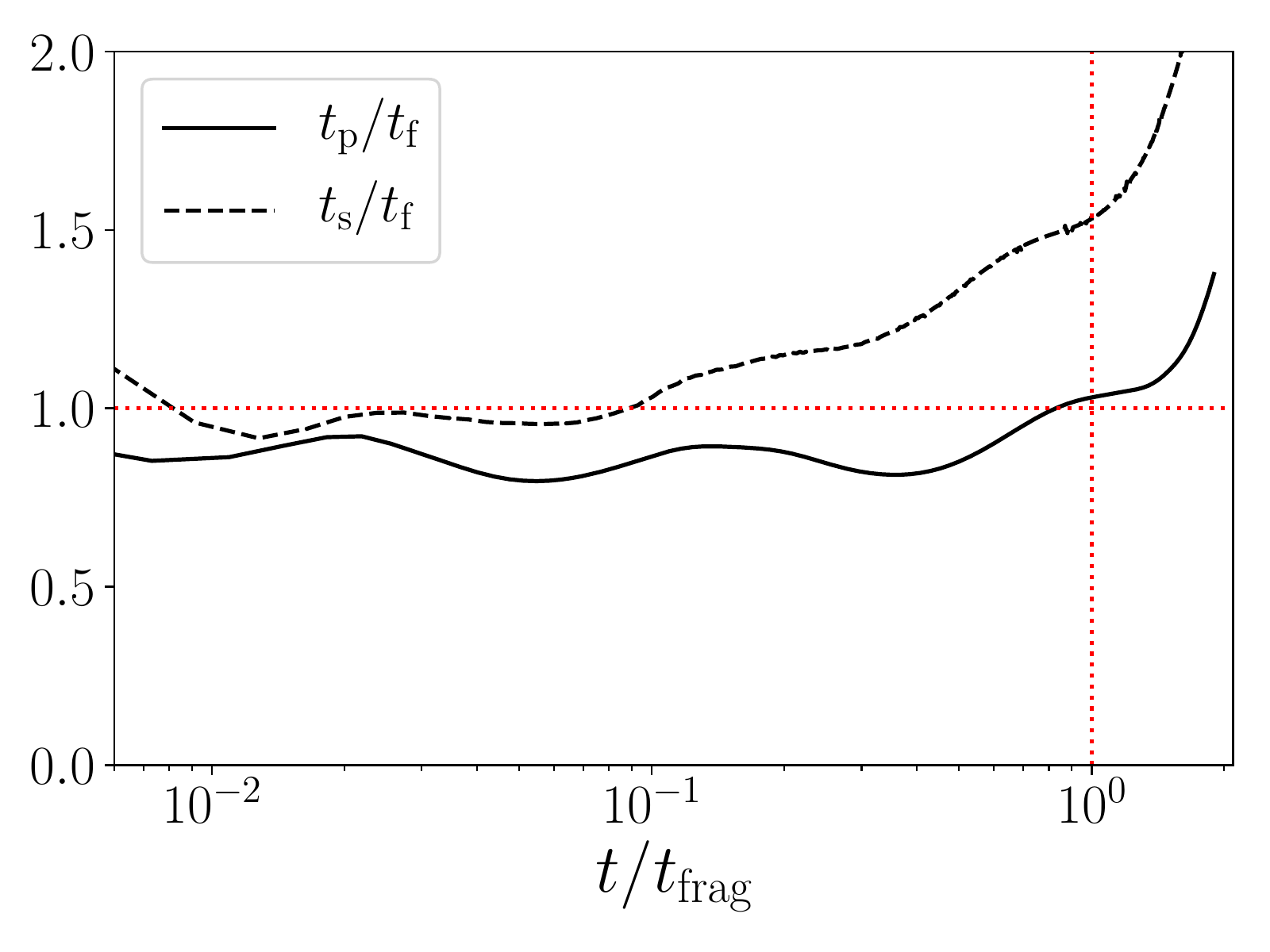}
    \caption{The solid black line is the ratio between thermal and free-fall time, the dashed black line between stretching and free-fall time. In red are shown the fragmentation time (vertical line) and the level at which the two considered time scales are equal (horizontal line).}
    \label{fig:tim_ra}
\end{figure}

The second important outcome of figure \ref{fig:tim_sca} is that the fragmentation occurs right after the free-fall time becomes shorter than the thermal time. This suggests that the fragmentation is a consequence of the pressure failing in balancing the stream self-gravity rather then an overcome of the black hole tidal forces or of the stream stretching by the gravitational collapse. This is better shown in figure \ref{fig:tim_ra}: the solid black line represents the ratio between the free-fall time and the thermal time while the dashed black line the ratio between free-fall time and stretching time. The two red dotted lines highlight the time at which fragmentation occur (the vertical one) and the line on which the two considered time scales are equals (horizontal one).

\subsection{Initial stellar rotation}

As explained section 1, a stream of gas debris obeying to a polytropic equation of state with $\gamma\gtrsim5/3$ is prone to fragmentation fueled by gravitational instabilities. The results of the previous section prove that also the $\gamma=5/3$ case is susceptible of fragmentation.

An interesting factor that can be taken into account and could potentially slow down the fragmentation process is the stellar rotation. In this paper, we will focus on a configuration where the stellar initial rotation shares direction and sense with the orbital angular momentum of the star itself.

We will not treat the case where the rotation is able to prevent the stellar tidal disruption. This occurs when the stellar rotation is sufficiently fast and its axis is parallel and opposite to the stellar orbital angular momentum, or within some tens degrees from this configuration. In this case the stream does not form or it is extremely faint \citep{sac19} and therefore is not relevant to the problem at hand.

If at least one component of the initial stellar rotation lies in the orbital plane there will be another force acting upon the debris stream: the centrifugal force inherited from the star. This force will add support against the collapse, counter-acting to some extent the self-gravity of the stream and making it harder for it to fragment.

In the case considered here, where the rotation axis is perpendicular to the orbital plane, no additional force is added to the picture described above, the only thing that gets modified is the stellar internal structure and mass-energy distribution, that gets wider. This is due to a swelling of the rotating star with respect to the non-rotating case. A wider mass-energy distribution would cause a faster stretching of the stream, thus slowing down the fragmentation process.

To better understand this configuration and test our prediction, we performed a numerical simulation, adding an initial stellar rotation ($\alpha=0.2$).

The simulations confirm our predictions about the role of stellar rotation aligned with the orbital angular momentum of the star: the time at which fragmentation occurs gets delayed by $\approx15\%$. The delay is caused by the bigger stretching that a wider initial mass-energy distribution provokes. 

\begin{figure}
	\includegraphics[width=\columnwidth]{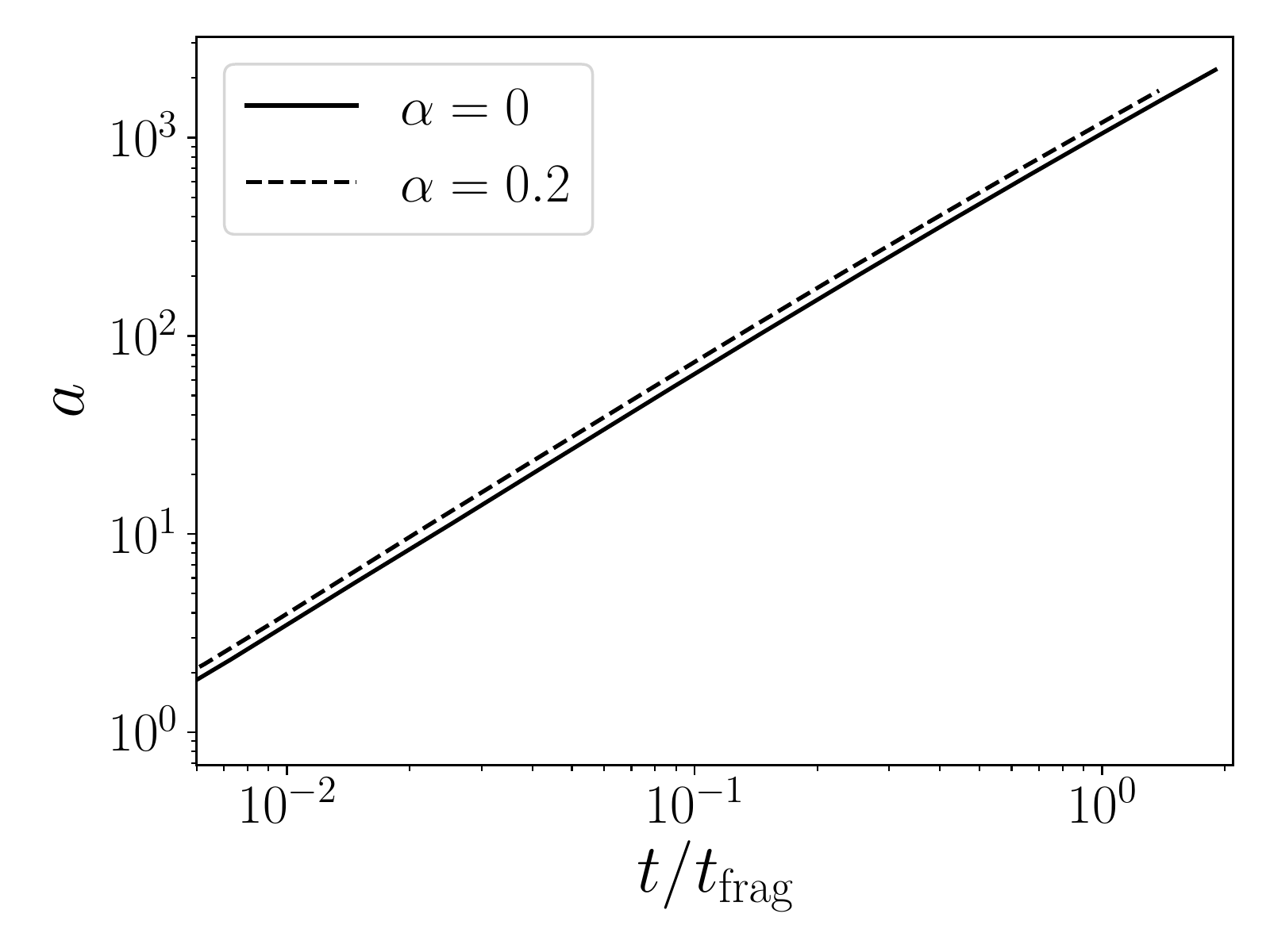}
    \caption{The plot shows the stretching parameter for the non-rotating (solid black line) and rotating case (dashed black line) as a function of time (normalized to the fragmentation time).}
    \label{fig:rot_stret}
\end{figure}

Figure \ref{fig:rot_stret} shows the stretching parameter for the non rotating (solid black line) and rotating case (dashed black line) as a function of time (at the fragmentation time specific of each configuration). One can notice how, in the rotating case, the stretching is bigger, but only by $\approx5\%$, this is however sufficient to produce the aforementioned amount of delay in the fragmentation time. This is due to the fact that the time scales have the same magnitude, thus a small changing in the value of one of the forces can lead to a significant difference in the evolutionary pattern of the stream.

\section{Analytical estimate of fragmentation condition}

We provide here a simple analytical framework to understand the physics behind the fragmentation process, described by our numerical simulations.

The formalism to study the equilibrium and the time evolution of cylinders of gas has been developed since the sixties of the past century to describe the filaments of gas found in gaseous star-forming regions (for a review, see i.e. \citealt{and14}). The standard case in the star-formation scenario is that a {\em contracting} cylinder increases its density while shrinking, accreting material from the parental cloud, until it becomes gravitationally unstable and fragments, thus forming smaller scale structures called cores. 
In the literature of this field several criteria have been found for the stability of polytropic cylinders of gas (see i.e, \citealt{ost64, inu92}). In particular, a fundamental condition that has to be satisfied in order to allow the gravitational collapse of filaments is that the linear mass-density of the filament should be larger than two times the square of the sound speed:
\begin{equation}
    \Lambda = \frac{{\rm d}M}{{\rm d}l}\gtrsim \frac{2c_{\rm s}^2}{G}
    \label{for:col_con}
\end{equation}
where $\Lambda={\rm d}M/{\rm d}l$ is the linear mass density computed as $\rho\upi H^2$.
We can translate mathematically this picture into our case just by inverting the sign of the adopted stretching. While in the star formation case the shrinking of the filaments enhances the tendency for collapse, in our case the opposite occurs, as the filament is stretched out to lower mean densities. Thus, in this case, the effect of the stretching is to dilute the fluctuations and can in principle prevent the formation of fragments if the expansion is sufficiently fast in comparison with the free-fall time.

We consider the stream as a gas cylinder (a similar approach has already been adopted by \citealt{cou19}). Furthermore we consider the cylinder to be in hydrostatic equilibrium along its transverse section (we will discuss later the limits of this assumption. However, for an analytical analysis of the stability of a polytropic cylinder of gas see i.e \citealt{toc15}). The transverse width of the stream is thus:
\begin{equation}
    H\simeq \frac{c_{\rm s}^2}{\sqrt{4\pi G\rho}}.\label{for:hydroeq}
\end{equation}
Finally, we assume the cylinder to be free-falling onto the black hole (this assumption too will be discussed later on).

We use cylindrical coordinates so that the cylinder length is $l(t)=l_0a(t)$ and its width is $H(t)=H_0b(t)$\footnote{Note that the analytic espression for the value of $H_0$ can be found i.e. in \citealt{ost64}.}, with the boundary conditions that $a(0)=b(0)=1$. 

\begin{figure}
	\includegraphics[width=\columnwidth]{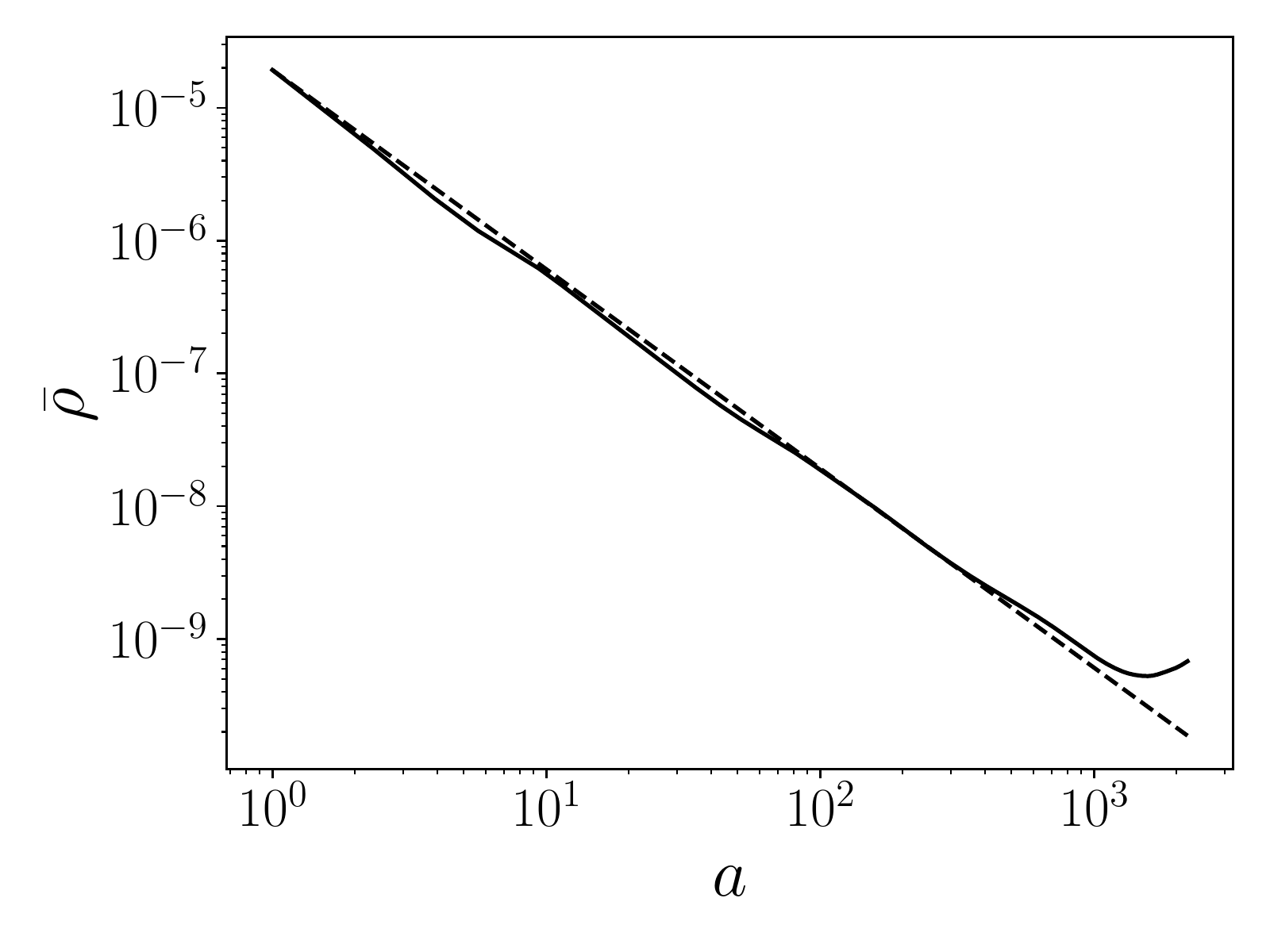}
    \caption{The plot shows the mean density of the stream (solid black line) as a function of the stretching parameter $a$. The dashed line indicates the $a^{-3/2}$ predicted behaviour for the $\gamma=5/3$ case covered by our simulation.}
    \label{fig:dens_stret}
\end{figure}

\begin{figure}
	\includegraphics[width=\columnwidth]{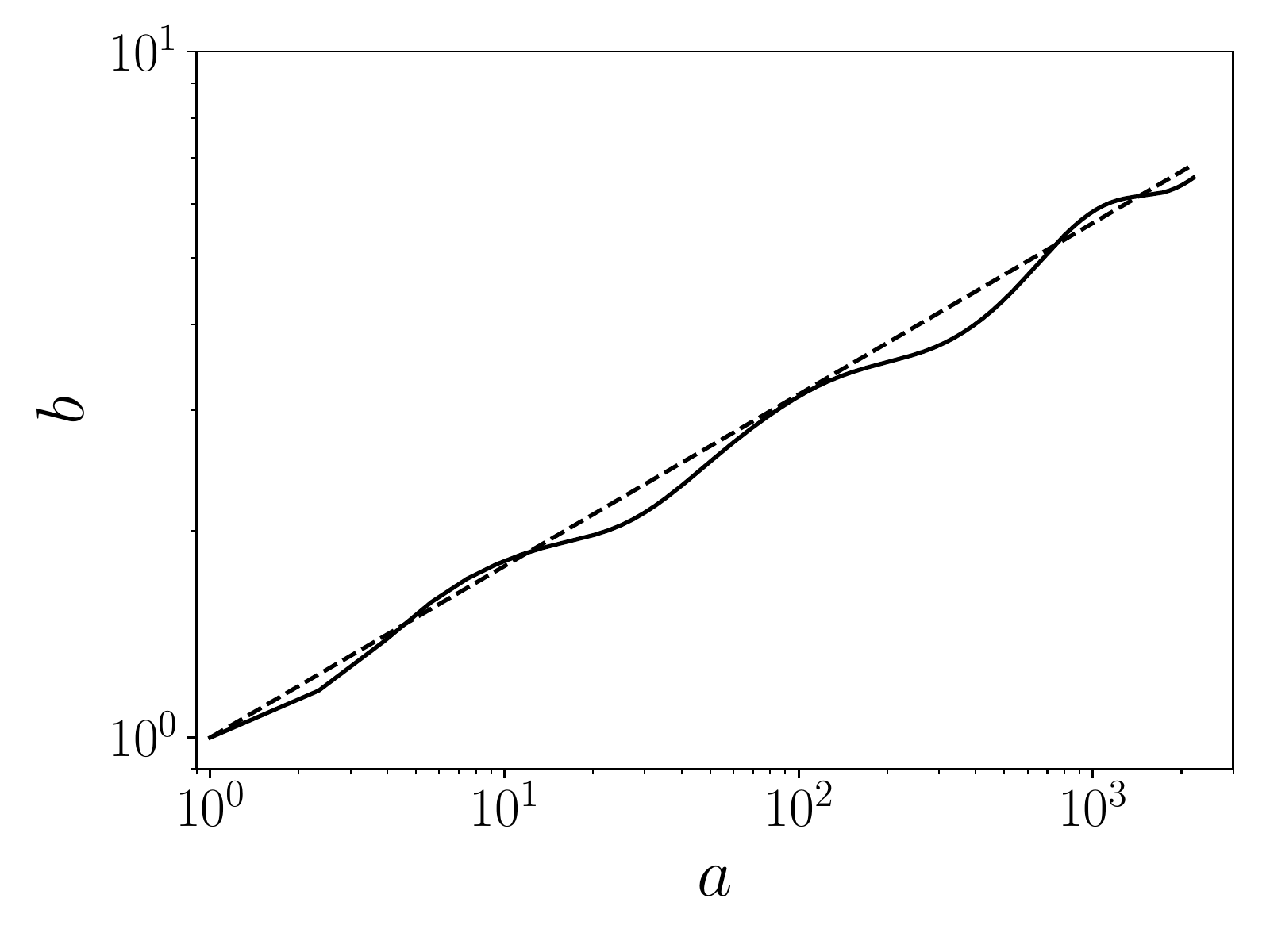}
    \caption{The plot shows the stretching parameter $b$ (solid black line) as a function of the stretching parameter $a$. The dashed line indicates the $a^{1/4}$ predicted behaviour for the $\gamma=5/3$ case covered by our simulation.}
    \label{fig:b_fac}
\end{figure}

The first step, in order to determine how all of the interesting quantities evolve, is to find a relation between the scaling factors $a$ and $b$ and link their time evolution to the evolution of the physical quantities of the stream. 
Mass conservation within the stream implies:
\begin{equation}
    H^2l\rho=const.,\label{for:masscons}
\end{equation}
which, combined with Eq. (\ref{for:hydroeq}), gives both the stream density as a function of $a$ and the polytropic index $\gamma$
\begin{equation}
    \rho\propto a^{1/(1-\gamma)},\label{for:dens}
\end{equation}
and a relation linking the scaling factors:
\begin{equation}
    b\propto a^{(2-\gamma)/(2\gamma-2)}.\label{for:bfac}
\end{equation}
For the $\gamma=5/3$ case we can compare this prediction with our simulation: figures \ref{fig:dens_stret} and \ref{fig:b_fac} show the behaviour of the mean density of the stream and the stretching parameter $b$, respectively (solid black lines) and the predicted dependency (dashed lines), as functions of the scaling parameter $a$. The expected behaviour is strikingly fulfilled for all of the stream evolution ($b\propto a^{1/4},\rho\propto a^{-3/2}$), prior to  fragmentation.

The time evolution of all the time-scales introduced in Section 3 is related to the time evolution of these key quantities, all expressed as a function of the scaling factor $a(t)$.
The knowledge of the time dependency of the scaling factor $a$ would allow us to derive all of the quantities as explicit functions of the time $t$ and the polytropic index $\gamma$. In the next section we derive an analytical solution for the evolution of the debris stream and therefore for the evolution of $a(t)$.\\

\subsection{Dynamics of the debris stream}

Here, we consider the dynamics of the debris stream after disruption as composed of test particles freely falling in the radial direction in the gravitational field of the black hole. This approach is the same as the one of \citet{cou16b}, who first considered the evolution of the structure of the debris stream in a semi-analytical way. They adopt the same simplification of radial freely falling particles and find that this is a very good approximation for the post-disruption hydrodynamical stream. \citet{cou16b} find approximate solutions for the position and velocity of the stream elements in the limit where the particles are close to the marginally bound orbit, that corresponds to the stream center of mass. Here, as described below, we generalize the solution of \citet{cou16b} and find an exact analytical solution for the whole of the stream. We then argue that it is the deviation from the ``close to marginally bound'' orbits that causes fragmentation.

The equation of motion of the debris is:
\begin{equation}
    \frac{1}{2}\left(\frac{{\rm d}r}{{\rm d}t}\right)^2=\frac{GM_\textup{h}}{r}+E,
\end{equation}
where bound, marginally bound, and unbound orbits are defined by $E<0$, $E=0$ and $E>0$, respectively. The solutions to these equations are well know since they are the standard Friedmann equations used in Cosmology to describe a matter dominated universe in the closed, flat and open case, respectively \citep{fri22}. For the marginally bound orbit (corresponding to a flat Universe) we have (see also eq. 8 in \citealt{cou16b})
\begin{equation}
    R(t)=\left(\frac{9}{2}GM_\textup{h}\right)^{1/3}t^{2/3},
\end{equation}
for the unbound debris (corresponding to an open Universe) we have a parametric solution:
\begin{equation}\label{eq:orb_unbound}
    r(\eta)=r_0(\cosh\eta-1), \qquad
    t(\eta)=t_0(\sinh\eta-\eta),
\end{equation}
while for the bound debris (corresponding to a closed Universe) we have:
\begin{equation}\label{eq:orb_bound}
    r(\eta)=r_0(1-\cos\eta), \qquad
    t(\eta)=t_0(\eta-\sin\eta),
\end{equation}
where $r_0=GM/|2E|$ and $t_0=GM/|2E|^{3/2}$ are the definition of a scale radius and a scale time, that depend on the energy of the debris $E$. These expressions give parametrically the exact positions of the debris elements across the whole stream (cf. the approximate positions given in Eq. 19 in \citealt{cou16b})

\begin{figure*}
	\includegraphics[width=\columnwidth]{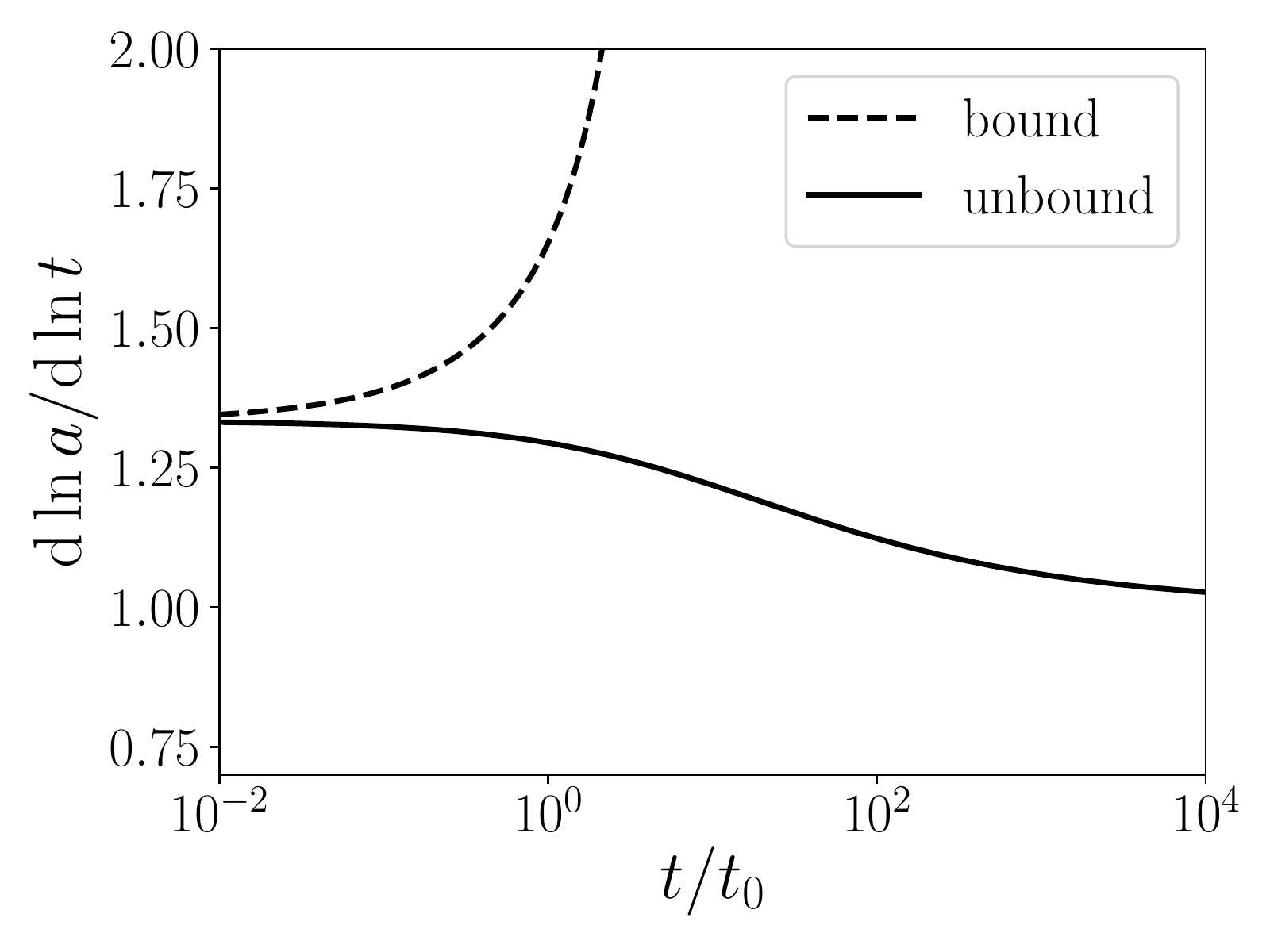}
	\includegraphics[width=\columnwidth]{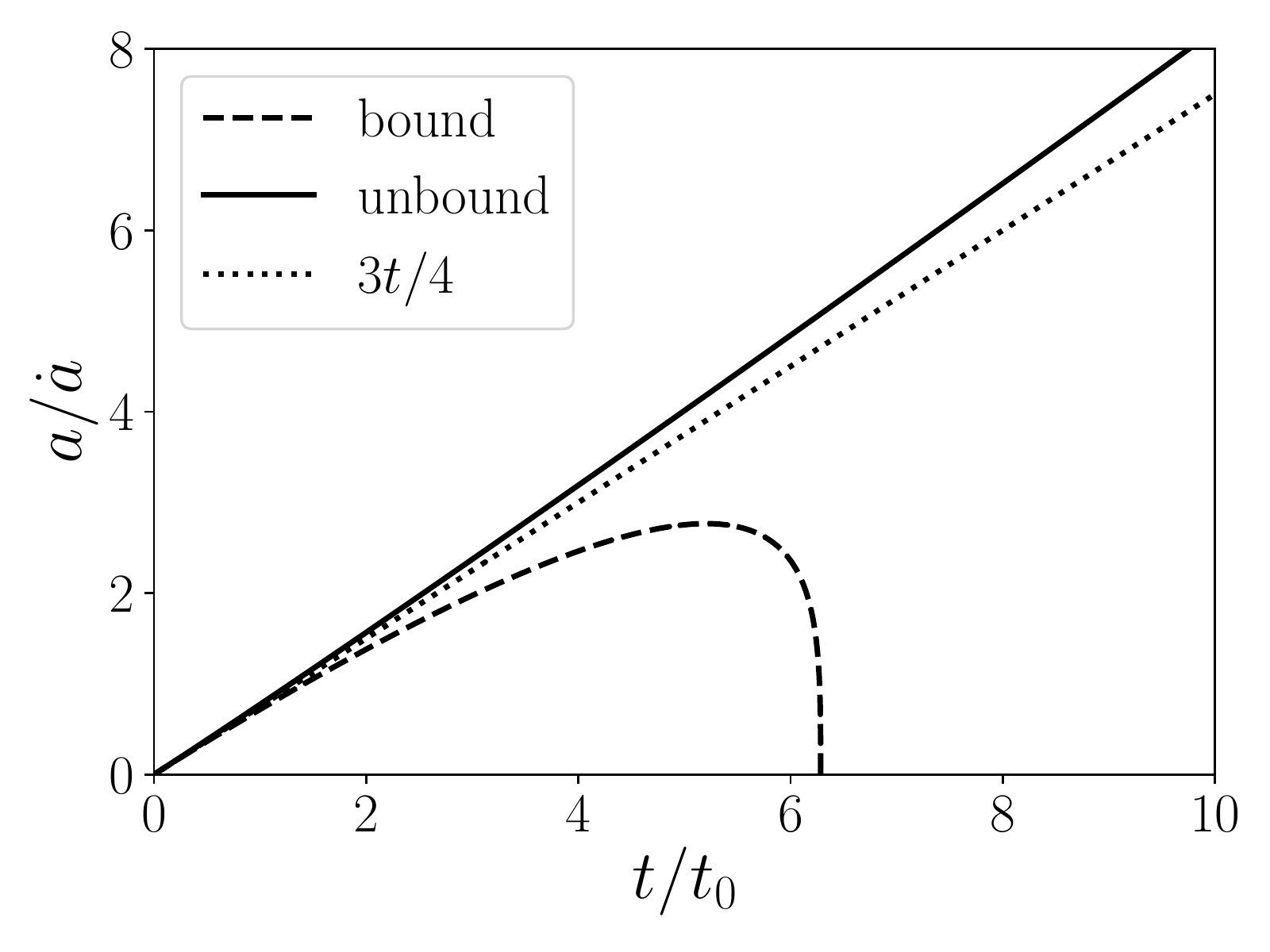}
    \caption{Left: logarithmic derivative of the stretching parameter from our analytical solutions for the bound and the unbound portion of the stream  as functions of time (normalized to the the scale time $t_0$). Right: evolution of the stretching time $a/\dot{a}$ for the bound and unbound debris, compared to the approximate value in the marginally bound case (dotted line).}
    \label{fig:stret_ana}
\end{figure*}

It is instructive to approximate the above solutions for small $\eta$, which corresponds to orbits close to the marginally bound case. 
This is obtained by expanding the hyperbolic and the trigonometric sine and cosine as:
\begin{equation}
    \cosh\eta-1 \sim \frac{\eta^2}{2}+\frac{\eta^4}{24}, \qquad
    \sinh\eta -\eta \sim \frac{\eta^3}{3},
\end{equation}
\begin{equation}
    1-\cos\eta \sim \frac{\eta^2}{2}-\frac{\eta^4}{24}, \qquad
    \eta-\sin\eta \sim \frac{\eta^3}{3}.
\end{equation}
In this limit, for both bound and unbound orbits, we obtain:
\begin{equation}
    \label{eq:pos_marginally}
    r(t)=R(t)\left(1+\frac{1}{3}\frac{ER(t)}{GM}\right).
\end{equation}
The relative position with respect to the center of mass of the stream is then 
\begin{equation}
l(t)=r(t)-R(t)= \frac{1}{3}\frac{E}{GM}R(t)^2\sim t^{4/3},   
\end{equation}
that implies that $a\propto t^{4/3}$ (cf. again \citealt{cou16b}, their eq. 16). Actually, our exact solution allows us to compute the evolution of $a(t)$ for the whole stream. Figure \ref{fig:stret_ana} (left panel) shows the logarithmic derivative of the stretching parameter $a(t)$ for the bound (dashed line) and the unbound (solid line) portion of the stream. One can see that initially both are close to the expected value of 4/3, while at late times (which occurs progressively earlier in physical time the farther we move away from the marginally bound orbit) the stretching of the unbound debris slows down, eventually reaching freely streaming orbits, where $a(t)\propto t$, while the bound debris are stretched faster, as the tide of the black hole increases. Similarly, one can compute the stretching timescale $t_{\rm s}=a/\dot{a}$, shown in Fig. \ref{fig:stret_ana} (right panel), which grows faster (slower) than $3t/4$ for the unbound (bound) portion of the stream. 

\citealt{cou16b} obtain their evolution by solving in a relatively complicated way the equation of motion of the debris, by introducing a parameter $\xi$, defined as:
\begin{equation}
    \xi=\sqrt{\frac{2GM_{\rm h}}{r^3}}t=\frac{2}{3}\left(\frac{R}{r}\right)^{3/2},
\end{equation}
which is a measure of how far from the marginally bound orbit a stream element is. They further make the reasonable but in principle not justified assumption that the stream velocity follows a self-similar solution:
\begin{equation}
    v_r=\sqrt{\frac{2GM_{\rm h}}{r}}f(\xi),
\end{equation}
and finally solve numerically for $f$ from an ordinary differential equation. They also provide an approximate solution for $f(\xi)$ in the form $f=1/\xi-1$, which they then use to compute the evolution of the stream far from the marginally bound orbit.  

Our approach allows us to bypass all this and obtain an analytical and exact solution for $f(\xi)$, at the same time allowing us to demonstrate that the radial velocity is indeed self-similar. 

Let us first consider the behaviour close to the marginally bound orbit, Eq. (\ref{eq:pos_marginally}). The radial velocity is readily obtained:
\begin{equation}
    v_r=\frac{\mbox{d}r}{\mbox{d}R}\frac{\mbox{d}R}{\mbox{d}t}= \sqrt{\frac{2GM_{\rm h}}{R}}\left(\frac{2r}{R}-1\right) =
    \sqrt{\frac{2GM_{\rm h}}{r}}\left[2\left(\frac{r}{R}\right)^{3/2}-\left(\frac{r}{R}\right)^{1/2}\right],
\end{equation}
and, recalling the definition of $\xi$, we obtain
\begin{equation}
    f(\xi)=\frac{4}{3\xi}-\left(\frac{2}{3\xi}\right)^{1/3},
    \label{eq:f_approx}
\end{equation}
which has the properties $f(2/3)=1$ and $f'(2/3)=-5/2$ (cf. \citealt{cou16b}). We thus see that, indeed, the \emph{ansatz} by \citet{cou16b} that the solution would be self-similar with respect to the variable $\xi$ is indeed correct, at least for particles close to the marginally bound orbit, as it only depends on how far does the particle lie with respect to the stream center of mass. We shall see in a moment that the self-similarity extends to the whole stream exactly. 

\begin{figure}
	\includegraphics[width=\columnwidth]{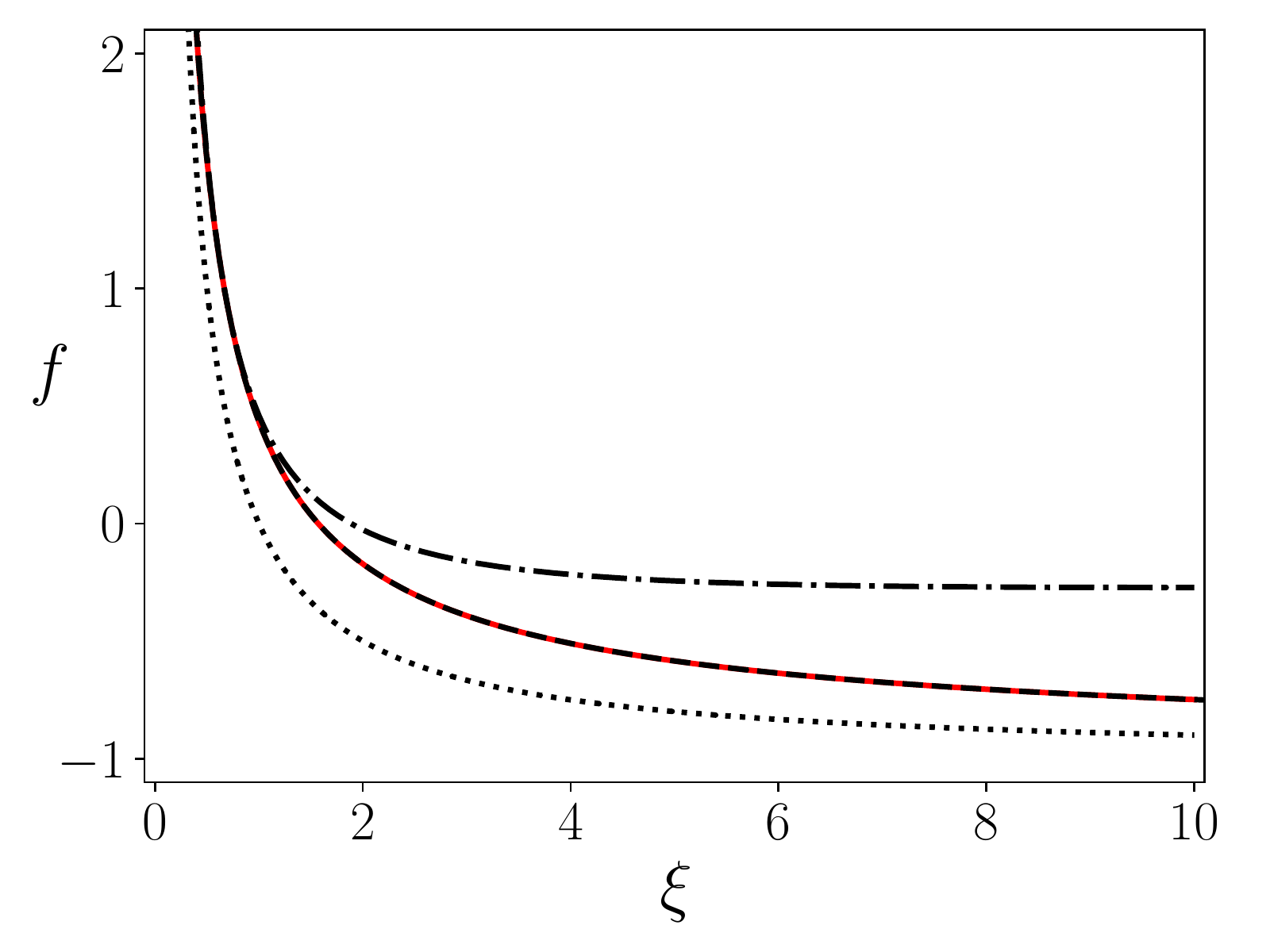}
    \caption{The dashed black line shows the analytical function $f$ superposed to the solid black line obtained solving numerically Eq. (4) by \citet{cou16b}. The dotted line and the dash-dotted one show respectively the approximated solutions proposed by \citet{cou16b} and by us.}
    \label{fig:self_sim}
\end{figure}
\begin{figure}
	\includegraphics[width=\columnwidth]{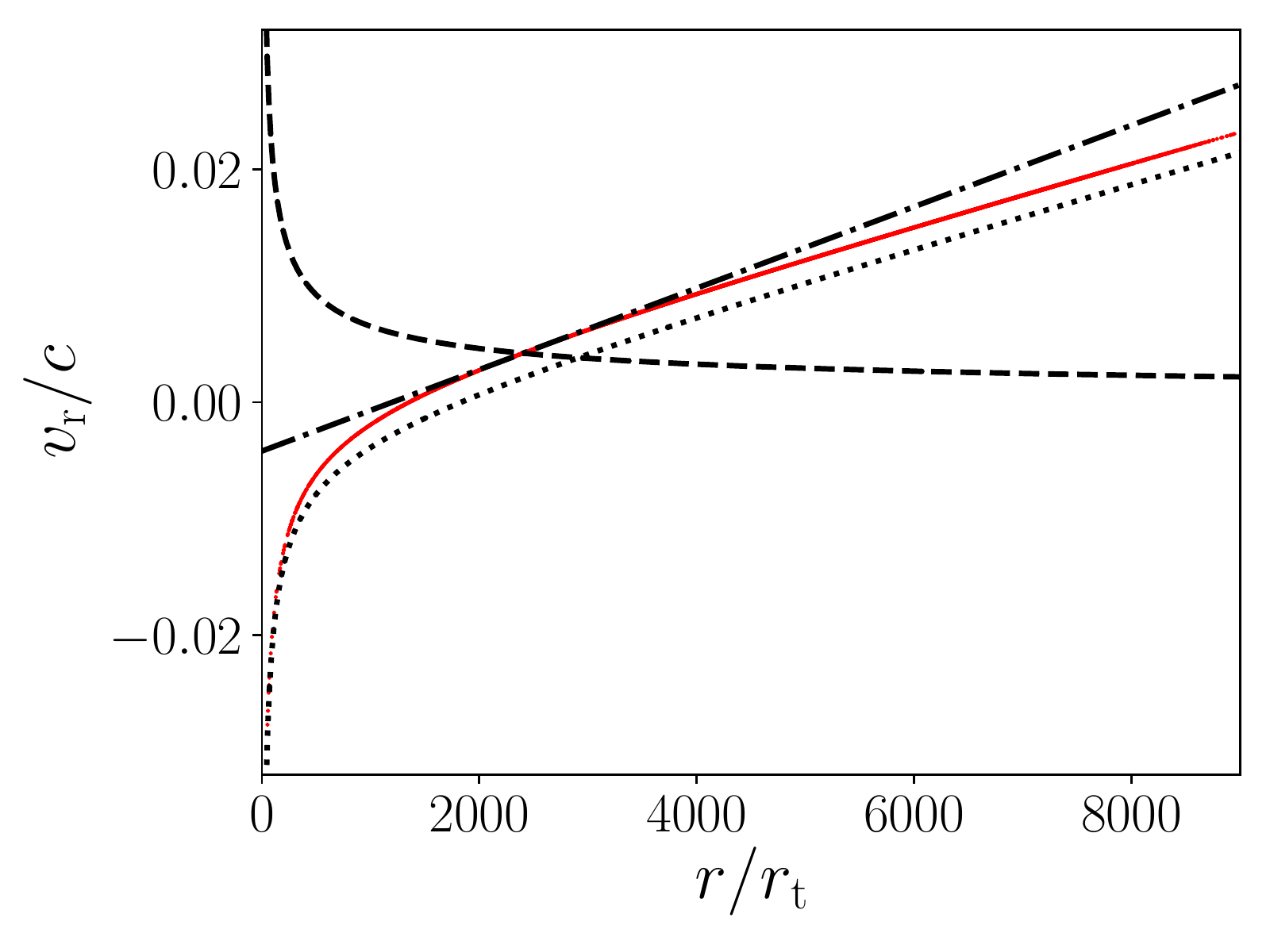}
    \caption{The red line shows the velocity distribution of the stream from our SPH simulation at $t\approx25t_{\rm min}$. The velocity distribution obtained using the  approximate formula proposed by us (which applies only close to the marginally bound debris) is shown with a dash-dotted line, while the approximate solution by \citet{cou16b} is shown with a dotted line. We do not plot here our exact solution because it coincides perfectly with the simulated one. We also plot the radial velocity profile of the marginally bound material (dashed line). The intersection of this line with the simulated one indicates the location of the marginally bound debris in the stream.}
    \label{fig:rad_vel}
\end{figure}

The function in Eq. (\ref{eq:f_approx}) approximates very well the full solution in the vicinity of the marginally bound orbit but fails farther from it. Actually, an exact and closed form analytical expression for $f(\xi)$ can be obtained by differentiating Eqs. \ref{eq:orb_unbound} and \ref{eq:orb_bound}:
\begin{equation}
    v_r=\frac{(\mathrm{d}r/\mathrm{d}\eta)}{(\mathrm{d}t/\mathrm{d}\eta)}.
\end{equation}
After some simple algebra, we can thus obtain $f(\xi)$ parametrically:
\begin{equation}
    f=\frac{\sin\eta}{\sqrt{2(1-\cos\eta)}}, \qquad
    \xi = \frac{\sqrt{2}(\eta-\sin\eta)}{(1-\cos\eta)^{3/2}},
    \label{eq:f_exact1}
\end{equation}
for the bound portion of the stream ($\xi>2/3$) and
\begin{equation}
    f=\frac{\sinh\eta}{\sqrt{2(\cosh\eta-1)}}, \qquad
    \xi = \frac{\sqrt{2}(\sinh\eta-\eta)}{(\cosh\eta-1)^{3/2}},
    \label{eq:f_exact2}
\end{equation}
for the unbound portion of the stream ($\xi<2/3$). In Figure \ref{fig:self_sim} we plot the analytical function $f(\xi)$ as described above (dashed black line), the numerically computed function based on solving eq. (4) in \citet{cou16b} (solid red line), and the two approximations: (i) the function $f=1/\xi-1$ proposed by \citet{cou16b} (dotted line) and (ii) the function we propose in Eq. (\ref{eq:f_approx}) above (dash-dotted line). As we can see, Eq. (\ref{eq:f_approx}) provides a better approximation (with respect to $f=1/\xi-1$) to the actual solution close to the marginally bound orbits, but fails to reproduce the asymptotic behaviour, especially for the bound portion of the stream ($\xi\gg 1$). On the contrary, the exact solution, Eqs. (\ref{eq:f_exact1}) and (\ref{eq:f_exact2}), does recover the numerically computed one. We also plot in Fig. \ref{fig:rad_vel} the resulting radial velocity at a given time $t\approx25\,t_{\rm min}$, where in red we show the distribution obtained from our SPH simulation and the various lines indicate the distribution obtained using our approximate solution close to the marginally bound debris (Eq. (\ref{eq:f_approx}), dash-dotted line) and the one from \citet{cou16b} approximated function described above (dotted line). We do not plot our exact solution because it coincides exactly with the numerical simulation. We also indicate, with a dashed line, the marginally bound orbit velocity profile. The intersection of this line with the simulation data marks the location of the marginally bound debris. Again, the general conclusion is that our exact solution is an excellent representation of the simulation data, that our approximate expression (Eq. (\ref{eq:f_approx}) is a good approximation close to the marginally bound debris but fails away from it, and that the \citet{cou16b} proposed function, while not approximating the actual solution at any point, gives a fair first order description of the asymptotic regimes.

We are now in a better position to evaluate the stability of the stream. We have seen that the stretching of the orbits initially proceeds as $t^{4/3}$, but then slows down in the unbound portion of the stream and accelerates in the bound portion. 

\subsection{Stream stability}

Now that we know the time dependency of the stream density and the scaling factors, it is straightforward to derive all of the other quantities and time-scales behaviours.

In Section 3 we introduced four time scales, each linked to one of the force acting on the stream. Here we will neglect the tidal time as this is much larger than the others and thus is not expected to play a role in the fragmentation process.

We have then the stretching time, the thermal one and finally the free-fall time. The first two stabilise the stream while the latter is ultimately responsible for fragmentation.

The stretching time dependence is the easiest to derive as this time scale is simply defined as $t_\textup{s}=a/\dot a$ and therefore
\begin{equation}
    t_\textup{s}\propto t,
\end{equation}
at least initially, bearing no dependence on the polytropic index.

We recall that the thermal time is defined as $t_\textup{p}=H/c_\textup{s}\propto 1/\sqrt{\rho}$, where we used the condition of hydrostatic balance in the transverse direction. Since $\rho\propto a^1/(1-\gamma)$ (Eq. \ref{for:dens})
\begin{equation}
    t_\textup{p}\propto a^{1/2(\gamma-1)}\propto
    \begin{cases}
        &t^{2/3(\gamma-1)}\qquad t\ll1\\
        &t^{1/2(\gamma-1)}\qquad t\to\infty\\
    \end{cases}.
\end{equation}

Finally, also the free-fall time scale, defined as $t_\textup{ff}=1/\sqrt{2\upi G\rho}$ is proportional to $1/\sqrt{\rho}$ and thus scales with time in the same way as the thermal time. Actually, the assumption of hydrostatic balance in the transverse direction implies that the stream is always close to marginal stability, according to the \citet{ost64} criterion, $\Lambda\sim c_{\rm s}^2/2G$ and thus fragmentation would ensue relatively easily. However, as long as the stream stretching occurs on the same timescale, this can act to stabilise it. 

It is therefore clear why the $\gamma=5/3$ case is the critical one: for this particular value of the polytropic index all of the relevant time scales share the very same time dependence, at least initially: $t_\textup{ff,sc,s}\propto t$, the stream stays therefore marginally stable for a long time. After some time however the time evolution of the thermal and free fall time scales slows down and they disentangle from the stretching time. This can be seen in figure \ref{fig:tim_sca}, and is due to the fact that at late times (when the stream starts to freely stream) they tend to be proportional to $t^{3/4}$ while the stretching time remains proportional to $t$.

In the case of more compact stars ($\gamma>5/3$) the stream stretching along its axis is less efficient than all of the others time scales and therefore the streams fragments much faster than in the $\gamma=5/3$ case. In all of the cases where $\gamma<5/3$ the stretching grows faster than the other forces and is therefore able to sustain the stream and prevent its fragmentation.

\begin{figure}
	\includegraphics[width=\columnwidth]{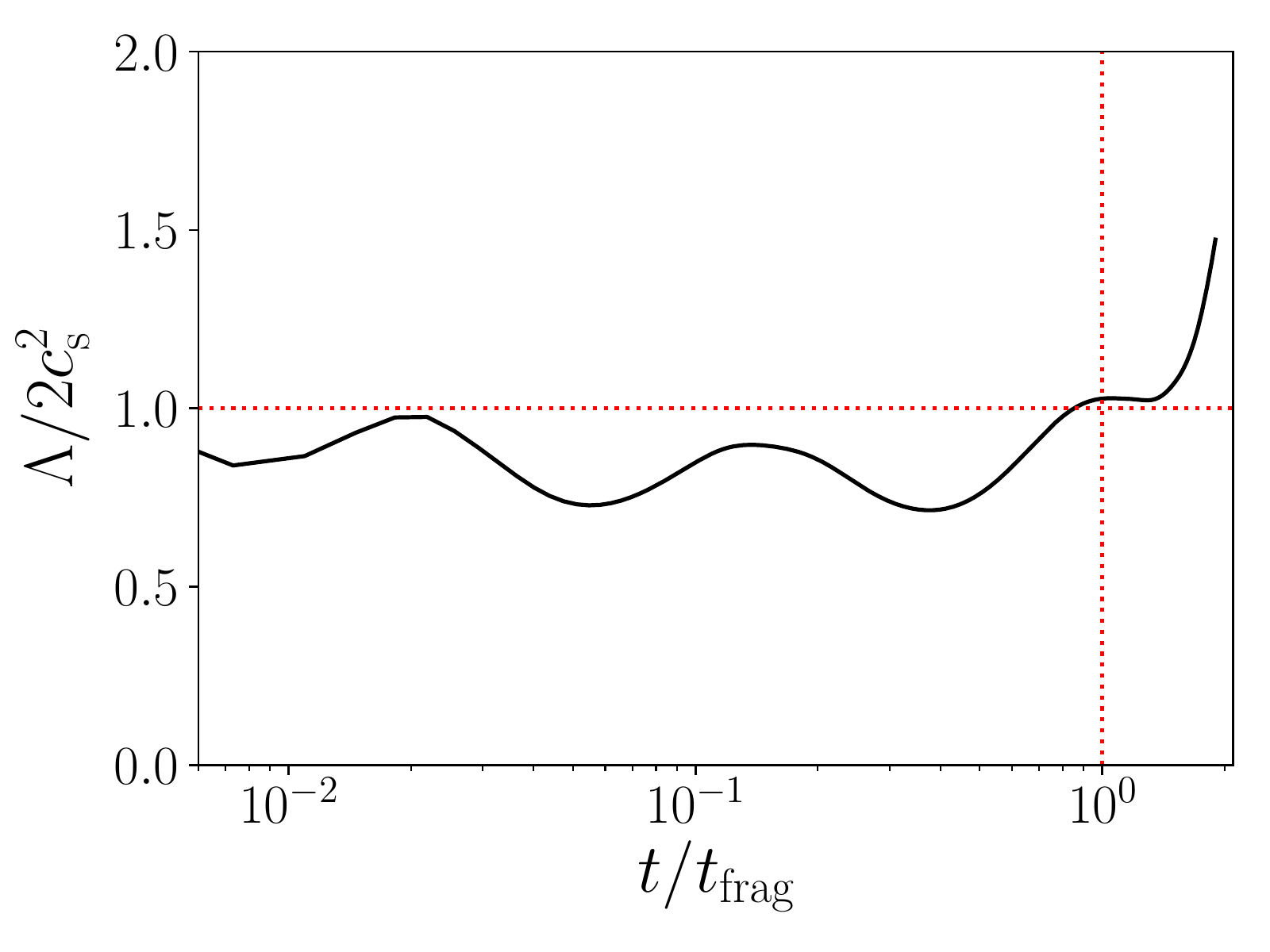}
    \caption{The solid black line shows the quantity $\Lambda/2c_\textup{s}^2$ that must be grater then 1 in order for a gas cylinder in hydrostatic equilibrium to become gravitationally unstable and fragment.}
    \label{fig:lin_den}
\end{figure}

Figure \ref{fig:lin_den} shows with a solid black line the quantity described in \eqref{for:col_con} while the two dotted red lines highlight again the fragmentation time and the level at which the condition is satisfied. We can see that the condition is almost satisfied during all of the stream evolution. It rapidly grows above unity right before the stream collapses, hence confirming that the stream can be considered in hydrostatic equilibrium for the part of its evolution we are interested in.\\

\begin{figure}
	\includegraphics[width=\columnwidth]{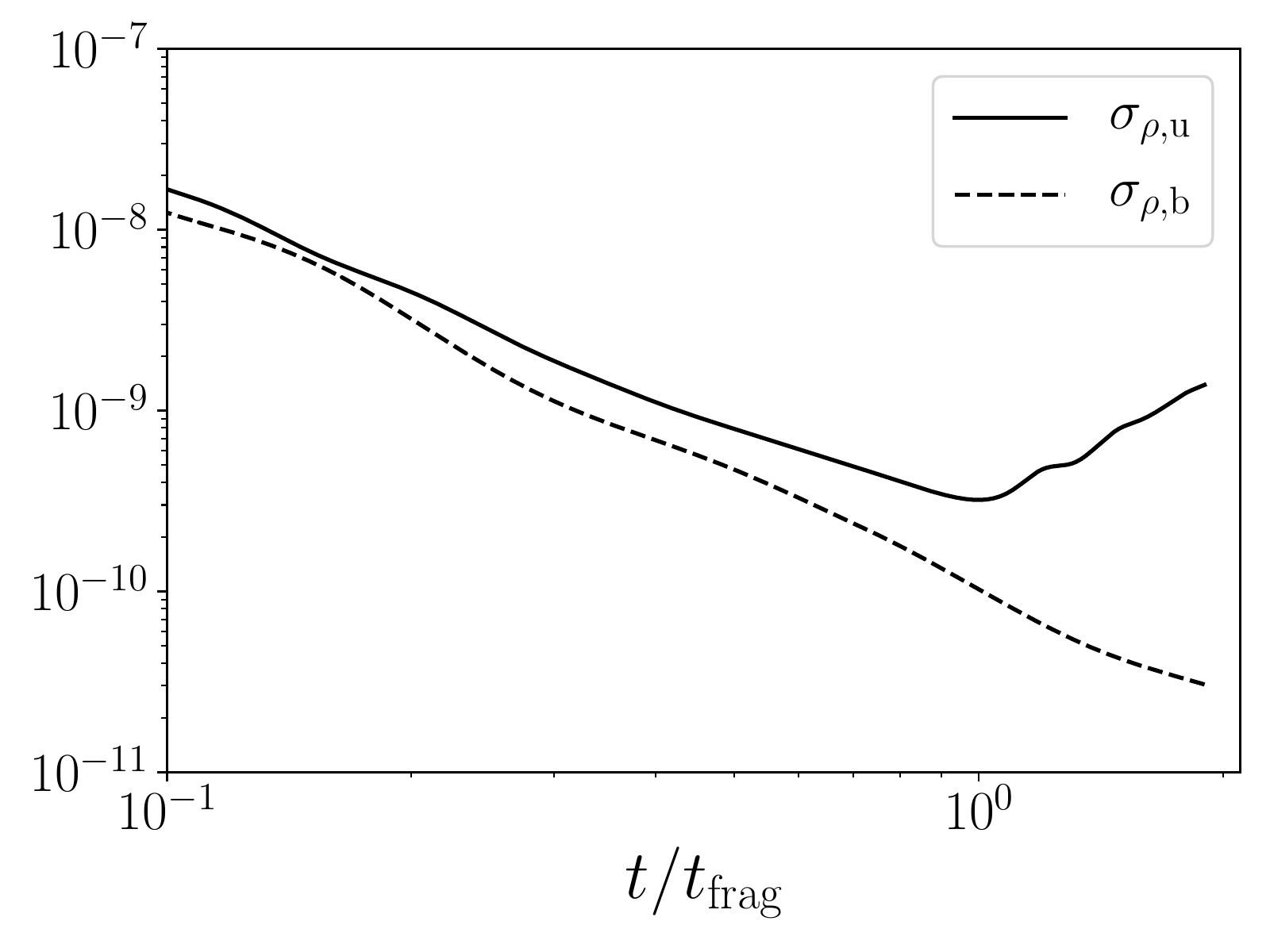}
    \caption{The plot shows the density fluctuations of the unbound (solid black line) and bound (dashed black line) part of the stream as functions of the fragmentation time of the unbound material.}
    \label{fig:den_fluc}
\end{figure}

All of the arguments presented above consider only the unbound part of the stream. Employing our analytical results we can however infer that in the bound part the stretching time falls rapidly below all of the other time-scale and thus should prevent the fragmentation of that portion of the stream. This is indeed what we observe in our simulations, too. Figure \ref{fig:den_fluc} shows the density fluctuations in the unbound (solid black line) and bound (dashed black line) part of the stream as a function of time. The absence of a turning point in the stream of bound material signals that little to no fragmentation is found in the bound debris before they enter the region in which they are considered to be accreted.

\section{Conclusions}

The pioneering work of \citet{cou15}, further developed by \citet{cou16a, cou16b}, studied, among other things, the gravitational stability of the debris streams forming after the disruption of a star by a SMBH. They found that a stream composed by a gas obeying a polytropic equation of state with polytropic index $\gamma\gtrsim5/3$ is susceptible to fragmentation. This fragmentation is caused by the onset of gravitational instabilities, that are able to overcome pressure forces within the gas and the tidal force of the black hole. The stream therefore would collapse under its self-gravity forming spheroidal blobs of gas. It is not however completely clear whether the critical $\gamma=5/3$ is subject to fragmentation or not. 
In order to confirm whether also the $\gamma=5/3$ case is affected by gravitational fragmentation, and to better understand the physic behind this process, we performed high resolution numerical simulation using a 3D SPH code. 

We found that a stream of gas debris resulting from the disruption of a star, modelled as a polytropic sphere with adiabatic index $\gamma=5/3$, is indeed prone to fragmentation. Through a convergence test we determined however that, for standard TDE parameters, fragmentation occurs only after more than 3 years since disruption, when the TDE has likely faded below observability.

We have also successfully described the process leading to stream fragmentation using an analytical approach that generalize the results of \citet{cou16b}, assuming that the stream can be modeled as a stretching cylinder and deriving a fragmentation condition that is more accurate than the ones previously suggested in the literature. 
In this picture, fragmentation is driven by the stream self-gravity and is resisted by pressure and by the stream stretching along its axis. For $\gamma=5/3$, as the stream expands, initially all the timescales associated with collapse, pressure and stretching grow with time at the same rate. However, at later times, we have demonstrated that the stretching time-scale is unable to keep up with the two others time-scales and fragmentation ensues rapidly. Conversely, for $\gamma > 5/3$, our argument predicts that the stretching time scale is always much longer than the pressure and self-gravity timescales, implying an increased tendency for collapse, as observed.

Further, our analysis predicts that the stretching time-scale should fall below all the others in the bound portion of the stream stabilizing it. This is indeed what we observe in our simulations: the bound material shows little to no fragmentation before entering the region where we consider it to be accreted.

While we were working on the present paper, we became aware of the recent work by \citet{cou20} on the general problem of the stability of a hydrostatic adiabatic self-gravitating filament. They find that the filament in unstable and propose that the very same instability is at the origin of stream fragmentation in TDE. The applicability of such analysis to TDE is not immediate, as already noted by \citet{cou20}, since (i) a TDE stream is not hydrostatic and (ii) the stream evolves significantly due to the presence of the tidal field of the black hole, that is not included in their analysis. Here, we propose instead that the origin of the fragmentation, as discussed above, lies in the slowing down of the stretching of the debris in the unbound portion of the stream.

\section*{Acknowledgements}

We thank an anonymous referee for very constructive criticism. We thank Eric Coughlin for very interesting discussions. 
We thank Daniele Galli for the priceless patience and help and Pierluigi Monaco for pointing out the analogy with Friedmann equations.
All the snapshots of our simulations were obtained using SPLASH \citep{pri07}.
This project and GL have received funding from the European Union's Horizon 2020 research and innovation program under the Marie Sklodowska-Curie grant agreement No 823823 (Dustbusters RISE project).
This work and CT have been supported by the project PRIN INAF 2016 The Cradle of Life - GENESIS-SKA (General Conditions in Early Planetary Systems for the rise of life with SKA).




\bibliographystyle{mnras}
\bibliography{biblio} 







\bsp	
\label{lastpage}
\end{document}